\documentclass{article}
\usepackage{amsmath}
\usepackage{amssymb}
\usepackage{comment}
\usepackage{algorithm,algpseudocode}
\usepackage{tikz}
\date{}
\newtheorem{theorem}{Theorem}
\newtheorem{lemma}{Lemma}
\newtheorem{claim}{Claim}
\newtheorem{definition}{Definition}

\newtheorem{question}{Open Question}

\newtheorem{proposition}{Proposition}

\newtheorem{corollary}{Corollary}
\title{Constant ratio FPT approximation of hypertree width parameters
for hypergraphs of bounded rank}
\author{Igor Razgon
\\ Department of Computer Science and Information Systems,\\ Birkbeck University of London \\ i.razgon@bbk.ac.uk}

\begin{document}
\maketitle
\begin{abstract}
We propose an algorithm whose input are parameters $k$ 
and $r$ and a hypergraph $H$ of rank at most $r$.
The algorithm either returns a tree decomposition of $H$ of generalized hypertree width at most $4k$
or 'NO'. In the latter case, it is guaranteed that the hypertree width of $H$ is greater than $k$. 
Most importantly, the runtime of the algorithm is \emph{FPT} in $k$ and $r$. 
The approach extends to fractional hypertree width with a slightly worse approximation ($4k+1$
instead of $4k$). We hope that the results of this paper will give rise to a new research direction
whose aim is design of FPT algorithms for computation and approximation of hypertree width
parameters for restricted classes of hypergraphs.  

\emph{Keywords: FPT algorithms, constant approximation, hypertree width, bounded rank.}
\end{abstract}
\section{Introduction}
















Hypertree width (including its generalized and fractional variants)  are well known structural parameters of hypergraphs.
A small value of a hypetree width parameter of the underlying hypergraph enables efficient evaluation of conjunctive queries 
and efficient \textsc{csp} solving. Therefore, an important computational task is to evaluate whether the considered parameter is small.
Formally, the input of such a task is a hypegraph and a (possibly non-integer) number $k>0$. The task is to test whether the
value of the considered parameter is at most $k$. 

A number of exact and approximation algorithms have been developed for this task,
for example \cite{HTWpaper,MarxFHWapprox,JACM21}.
However, all the algorithms we are aware of are \emph{\textsc{xp}} algorithms: their runtime is in the
form $O(n^{f(k)})$. To put it differently: these algorithms are polynomial for every fixed $k$
but the degree of the polynomial grows as $k$ grows. It is thus a natural question whether 
fixed-parameter tractable (\textsc{fpt}) algorithms are possible for this task. An \textsc{fpt} algorithm has 
runtime of the form $O(f(k) \cdot n^c)$ where $c$ is a constant or to put it differently, its runtime is  $O(n^c)$
for every fixed $k$.
As compared with \textsc{xp} algorithms, the degree of the polynomial in the runtime does not grow with the growth of $k$.
If we do not put any restrictions on the input then \textsc{fpt} computation of hypertree width parameters 
is not possible unless a widely believed complexity theoretical
conjecture fails. In particular, computation of hypertree width is W[2]-hard \cite{GottlobGMSS05}
generalized hypertree width (\textsc{ghw}) is \textsc{np}-hard to test for $k \geq 3$ 
\cite{2009gottlob}  and fractional hypertree width \textsc{fhw} is \textsc{np}-hard to test for $k \geq 2$ \cite{JACM21}.
We are not aware of results explicitly ruling out efficient \textsc{fpt} approximation for hypertree width but, taking into account 
that hypertree width evaluation can be considered as generalization of Set Cover problem and \textsc{fpt} inapproximability 
results \cite{Domsetfptnonapprox,Setcoverfptinapprox}, the perspective of a good \textsc{fpt} approximation for \textsc{ghw} or \textsc{fhw}  looks equally hopeless. We conclude that \textsc{fpt} algorithms might be possible only for restricted classes of hypergraphs.

The next natural question is what restricted classes to consider. Taking into account the connection
with Set Cover mentioned in the previous paragraph, it looks natural to consider a class for which Set 
Cover is known to be \textsc{fpt}.  To the best of our knowledge the most general class for which Set Cover is known
to be \textsc{fpt} are hypergraphs with bounded multiple intersection \cite{domsetnolargebiclique} 
\footnote{The result is stated for the Dominating Set problem, can be adapted to Set Cover with a minor tweaking}.
To put it differently, a class ${\bf H}$ of hypergraphs has a bounded multiple intersection if there is a constant
$c$ with such that for each $H \in {\bf H}$, the incidence graph of $H$ does not have $K_{c,c}$ as a subgraph. 
It is also interesting to note that such hypergraphs admit \textsc{xp} algorithm for evaluation of both generalizaed and
fractional hypeprtree width \cite{JACM21}.
We conclude that it makes sense to consider \textsc{fpt} algorithms for computing hypertree width parameters 
only for classes of hypergraphs with bounded multiple intersection.

In this paper we make the first step towards this direction by developing fixed parameter constant
approximation algorithms for computation of generalized and fractional hypertree width for hypergraphs
of bounded rank. In particular, we demonstrate the following.
\begin{enumerate}
\item There is an \textsc{fpt} algorithm parameterized by $k$ and $r$ whose input is a hypergraph of rank at most  $r$.
The algorithm returns either a tree decomposition of $H$ of \textsc{ghw} at most $4k$ or 'NO'. In the latter 
case it is guaranteed that the \textsc{ghw} of $H$ is greater than $k$. By allowing the approximation factor
to be increased to $6$, it is possible to remove the $|V(H)|$ factor from the runtime of the algorithm. 
\item There is an \textsc{fpt} algorithm parameterized by $k$ and $r$ whose input is a hypergraph of rank at most  $r$.
The algorithm returns either a tree decomposition of $H$ of \textsc{fhw} at most $4k+1$ or 'NO'. In the latter 
case it is guaranteed that the \textsc{fhw} of $H$ is greater than $k$.
\end{enumerate}

It is important to note that because of the rank assumption \textsc{ghw} or \textsc{fhw} of $H$ at 
most $k$ implies that the treewidth of the Gaifman graph $G$ of $H$ is at most $k\cdot r$.
Therefore, testing whether treewidth of $G$ is at most $k\cdot r$ provides an $O(r)$ approximation
for both \textsc{ghw} and \textsc{fhw}. The approximation rate of the algorithms proposed in this paper 
is constant, \emph{not} dependent on the rank, even though $r$ does appear in the runtime as the
extra parameter.

Rather than using a treewidth computation algorithm as a blackbox, we use the ideas developed
for the purpose of treewidth computation. In particular, in order to design the algorithms 
proposed in this paper, we adapt a well known approach for treewidth computation that is based
on successive computation of balanced separators \cite{GraphMinors13, ReedTWD}.
In the context of hypertree width computation, this framework was used for an
\textsc{xp}-time cubic approximation of \textsc{fhw} \cite{MarxFHWapprox}.
  
As a result we obtain a \emph{polynomial time} algorithmic framework that needs two 
oracles. 
For the purpose of \textsc{ghw} computation, 
one oracle gets as input a hypergraph $H$, a subset $S$ of its vertices and a parameter $p$ 
and tests whether the edge cover number of $S$ is at most $p$. 
The other function gets the same input under an extra assumption that the 
edge cover number of $S$ is $O(p)$ and tests whether $S$ has a balanced separator of  
of edge cover number at most $p$. The \textsc{fhw} is computed by the same framework using 
fractional versions of these oracles. The framework does not put any restrictions on
the input hypergraph. In other words, for any class of graphs admitting \textsc{fpt} algorithms for the oracles,
their substitution into the framework results in an \textsc{fpt} algorithm for the respective hypertree width 
parameter. We thus believe that the framework can be used for obtaining \textsc{fpt} algorithms for more
general classes of hypergraphs.

We obtain \textsc{fpt} algorithms for hypergraphs of bounded rank by designing \textsc{fpt} algorithms for 
these oracles. In fact, since input sets for the oracles have bounded edge cover number,
they also have bounded size. Therefore, their edge cover numbers can be tested in a brute-force
manner. To efficiently test existence of balanced separators, we obtain a result that might
be of an independent interest. In particular, we consider a problem whose input is a hypergraph
$H$, two distinct vertices $s$ and $t$ of $H$ and a parameter $k$. The task is to test existence
of an $s,t$-separator whose edge cover number is at most $k$. We show that this problem is \textsc{fpt}
parameterized by $k$ and the rank $r$ of $H$. The same holds for the fractional case where $k$
may be real, so the first parameter is $\lceil k \rceil$. To establish the fixed-parameter tractability,
we use the Treewidth Reduction Theorem \cite{TALG}. To the best of our knowledge, it is the first
time this approach is used for hypergraphs.

The proposed results demonstrate that an approach to treewidth approximation
can be successfully adapted for \textsc{fpt} approximation of hypertree width parameters.
It is thus plausible that such adaptation is possible for other treewidth algorithms.
We provide a more detailed exploration of this connection in the Conclusion section
where we discuss possible directions of further research.

The rest of the paper is organized as follows.
Section 2 provides the necessary background. The \textsc{fpt} approximations
for \textsc{ghw} and \textsc{fhw} are described in Sections 3 and 4 respectively.
Section 5 discusses further research.

\section{Preliminaries}
A \emph{hypergraph} $H$ is a pair $(V,E)$ (also referred as $V(H)$ and $E(H)$)
where $V(H)$ is the set of \emph{vertices} of $H$ and $E(H)$ is a family of 
subsets of $V(H)$ that are called \emph{hyperedges}.
The \emph{rank} of $H$ is the largest size of its hyperedge.
We assume that the hypergraphs $H$ considered in this paper do not contain isolated vertices
or, to put it differently,
$\bigcup_{e \in E(H)} e=V(H)$. 

$H$ is a \emph{graph} if all the hyperedges are of size exactly $2$.
When we refer to graphs, we use the standard terminology as e.g. in \cite{Diestel3}.
This terminology (mostly) naturally extends to hypergraphs.
We discuss several notions especially important for this paper. 

Let $U \subseteq V(H)$. The sub-hypergraph $H[U]$ of $H$ \emph{induced}
by $U$ has the set of vertices $U$ and the set of hyperedges
$\{e \cap U| e \in E(H), e \cap U \neq \emptyset\}$. 
The hypergraph $H \setminus U$ is $H[V(H) \setminus U]$.
Notice that this definition diverges from the notion of an induced subgraph
in that, in the induced sub-hypergraph, there may be hyperedges  of size $1$
while their preimages in the original hypergraph are of size greater than $1$. 
However, such a definition is very handy in that it ensures that the absence
of isolated vertices is preserved in the induced subgraphs and also, 
as we will see later, that 
it preserves edge cover numbers for subsets of $U$.

A \emph{path}  of $H$ is a sequence $v_1, \dots, v_r$ 
of vertices so that for every two consecutive vertices $v_i,v_{i+1}$
there is $e \in E(H)$ containing both $v_i$ and $v_{i+1}$. 
A \emph{connected component} of $H$ is a maximal subset $U$ of $V(H)$
so that every two vertices of $U$ are connected by a path. 
Sometimes we slightly abuse the notation and identify $U$ with $H[U]$. 

Let $s_1,\dots, s_q \in V(H)$, $q \geq 2$. An $s_1, \dots, s_q$-separator also referred to as
$s_1, \dots, s_q$-cut or a multicut for $s_1, \dots, s_q$ is $X \subseteq V(H) \setminus \{s_1, \dots, s_q\}$
such that in $H \setminus X$ there is no path between $s_i$ and $s_j$ for each $1 \leq i \neq j \leq q$ .
This notation naturally extends to sets of vertices. In particular, let $S_1,S_2$ be two disjoint subsets
of $V(H)$. Then an $S_1,S_2$-cut $X$ is a subset of $V(H) \setminus (S_1 \cup S_2)$ such that
im $H \setminus X$ there is no path between a vertex of $S_1$ and a vertex of $S_2$.

Let $E$ be a family of sets. We denote $\bigcup_{e \in E} e$ by $\bigcup E$.
Let $U \subseteq V(H)$, $E_0 \subseteq E(H)$. We say that $E_0$ \emph{covers}
$U$ if $U \subseteq \bigcup E_0$. The smallest size of a subset of $E(H)$ covering
$U$ is called the \emph{edge cover number} of $U$ and denoted by $\rho_H(U)$
(the subscript may be omitted if clear from the context).
Also, we denote $\rho(V(H))$ by $\rho(H)$. 

Next, we define the \emph{fractional edge cover number}.
Let $\gamma: E \rightarrow [0,1]$ be a function.
Then $\gamma$ naturally extends to $V(H)$ 
by setting $\gamma(v)=\sum_{v \in e} \gamma(e)$. 
Then $\gamma$ is a fractional edge cover for $U \subseteq V(H)$
if for each $u \in U$, $\gamma(u) \geq 1$.
The \emph{weight} of $\gamma$ is $\sum_{e \in E(H)} \gamma(e)$. 
The fractional edge cover number of $U$ is the smallest weight of a fractional
edge cover of $U$, it is denoted by $\rho^*_H(U)$, again the subscript can be omitted if clear 
from the context, and $\rho^*(V(H))$ is denoted by $\rho^*(H)$.


\begin{proposition} \label{rhopreserve}
Let $X \subseteq U \subseteq V(H)$.
Then $\rho_H(X)=\rho_{H[U]}(X)$ and $\rho^*_H(X)=\rho^*_{H[u]}(X)$. 
\end{proposition}

{\bf Proof.}
Let $E_U$ be the set of hyperedges of $H$ having a non-empty intersection with $U$. 
Let $g_U$ e the function with domain $E_U$ such that for each $e \in E_U$,
$g_U(e)=e \cap U$. 
Clearly, $E(H[U])=\{g_U(e)|e \in E_U\}$.

Let $\rho_H(X)=q$. This means that there are hyperedges $e_1, \dots e_q \in E_U$
such that $X \subseteq e_1 \cup \dots \cup e_q$.
Clearly, as $X \subseteq U$, it follows that
$X \subseteq g_U(e_1) \cup \dots \cup g_U(e_q)$.
That is, $\rho_{H[U]} \leq \rho_H(U)$. 

Conversely, let $\rho_{H[U]}(X)=q$ and let
$e'_1, \dots, e'_q \in E(H[U])$ be hyperedges covering $X$. 
Let $e_i$ be an arbitrary element of $g_U^{-1}(e'_i)$ for $1 \leq i \leq q$.
As each $e'_i$ s a subset of $e_i$, we conclude that $e_1, \dots, e_q$ 
cover $X$  thus demonstrating that $\rho_H(X) \leq \rho_{[U]}(X)$.

For the fractional case, 
let $\gamma:E(H) \rightarrow [0,1]$ be a witnessing function for $\rho^*_H(X)$. 
Note that for each $e \subseteq V(H) \setminus U$
$\gamma(e)=0$. Indeed, otherwise, we can set the weight of $e$ to zero still
covering $X$ with a smaller weight, a contradiction.

We define the function $\gamma_U: E(H[U]) \rightarrow [0,1]$ as follows.
For each $e' \in E(H[U])$, $\gamma_U(e')=min(1, \sum_{e \in \gamma_U^{-1}(e')} \gamma(e))$.  
Let us demonstrate that for each $v \in X$, $\gamma_U(x) \geq 1$.
If there is $e' \in E(H[U])$ containing $v$ such that $\gamma_U(e')=1$, we are done. 
Otherwise, $\gamma_U(x)=\sum_{x \in e} \gamma(e)=\gamma(x) \geq 1$.
It is also clear from the description that the weight of $\gamma_U$ does not exceed the weight of $\gamma$.
We thus conclude that $\rho^*_{H[U]}(X) \leq \rho^*_H(X)$. 

Conversely, let $\gamma_U: E(H[U]) \rightarrow [0,1]$ be a witnessing function
for $\rho^*_{H[U]}(X)$. We define a function $\gamma:E(H) \rightarrow [0,1]$ as follows.
For each $e' \in E(H[U])$, pick an arbitrary $e \in g_U^{-1}(e')$ and set $\gamma(e)=\gamma_U(e')$
For the edges $e$ of $E(H)$ whose weight has not been set this way, let $\gamma(e)=0$.  
It follows from the construction that the weight of $\gamma$ is the same as the weight 
of $\gamma_U$. For the coverage of $X$, let $v \in X$ and let $e'_1, \dots e'_r$ be the hyperedges
of $E(H[U])$ covering $v$. Then $\sum_{i=1}^r \gamma_U(e') \geq 1$. Let $e_1, \dots, e_r$ be the corresponding 
hyperedges for $e'_1, \dots, e'_r$ so that each $e_i \in g_U(e'_i)$ and $\gamma(e_i)=
\gamma_U(e'_i)$. As $e'_i \subseteq e_i$,
we conclude that $e_1, \dots, e_i$ provide a cover for $X$. 
Thus $\rho_H(X) \leq \rho_{H[U]}(X)$.
$\blacksquare$

The notion of a tree decomposition for graphs easily extends to hypergraphs.
In particular, a tree decomposition of $H$ is a pair $(T,{\bf B})$ where
$T$ is a tree and ${\bf B}: V(T) \rightarrow 2^{V(H)}$ that satisfies the following
axioms: (i) union, that is $\bigcup_{x \in V(T)} {\bf B}(x)=V(H)$ (ii) containement,
that is for each $e \in E(H)$, there is $x \in V(T)$ such that $e \subseteq {\bf B}(x)$
and (iii) \emph{connectedness}. The last property means that for each $u \in V(H)$
$T[X_u]$ is connected where $X_u$ is the set of all $x \in V(T)$ such that $u \in {\bf B}(x)$. 

We extend the function ${\bf B}$ from elements of $V(T)$ to subsets 
of $V(T)$ as follows: for $X' \subseteq V(T)$, ${\bf B}(X')=\bigcup_{x \in X} {\bf B}(x)$. 
Also, let $T'$ be a subgraph of $T$. Then we let ${\bf B}(T')={\bf B}(V(T'))$.  

The following statement will be important for our further reasoning. 
\begin{proposition} \label{decompsep}
Let $(T,{\bf B})$ be a tree decomposition of a hypergraph $H$.
Let $x \in V(T)$. Let $V'$ be a connected component of $T \setminus x$.
Then ${\bf B}(V') \setminus {\bf B}(x)$ is a union of connected components
of $H \setminus {\bf B}(x)$. To put it differently,  
let $V'$ and $V''$ be two distinct connected components of $T \setminus x$.
Then no hyperedge of $H$ has a non-empty intersection with both
${\bf B}(V') \setminus {\bf B}(x)$ and ${\bf B}(V'') \setminus {\bf B}(x)$. 
\end{proposition} 

{\bf Proof.}
Let $V',V''$ be two distinct connected components of $H \setminus {\bf B}(x)$. 
Suppose there is $e \in E(H)$ having a non-empty intersection with both
${\bf B}(V') \setminus {\bf B}(x)$ and ${\bf B}(V'') \setminus {\bf B}(x)$. 
By the containement property, there is $x' \in V(T)$ such that $e \subseteq {\bf B}(x')$.
If $x=x'$ then both  ${\bf B}(V') \setminus {\bf B}(x)$ and ${\bf B}(V'') \setminus {\bf B}(x)$
have the empty intersection with $e$, a contradiction. 
Assume w.l.o.g. that $x' \in V'$. 
In particular, this means that $e \subseteq {\bf B}(V')$.
It is a standard fact about tree decompositions, easily following from the connectedness property
that ${\bf B}(V') \cap {\bf B}(V'') \subseteq {\bf B}(x)$. In particular, this means that 
$({\bf B}(V') \cap e)  \cap ({\bf B}(V'') \cap e)) \subseteq {\bf B}(x)$.
Since, by assumption, ${\bf B}(V') \cap e=e$, we conclude that
${\bf B}(V'') \cap e \subseteq {\bf B}(x)$ implying that 
${\bf B}(V'') \setminus {\bf B}(x)$ does not intersect with $e$, again causing a contradiction.
$\blacksquare$

Proposition \ref{decompsep} has the following important consequence 
for assessment of edge cover numbers.

\begin{corollary} \label{decomprho}
Let $(T,{\bf B})$ be a tree decomposition of a hypergraph $H$. 
Let $x \in V(T)$ and let $V'_1, \dots, V'_r$ be the connected components
of $T \setminus x$. 
Let $U_1, \dots U_r$ be respective subsest of $V'_1 \setminus {\bf B}(x), \dots V'_r \setminus {\bf B}(x)$. 
Then $\rho(\bigcup_{i=1}^r U_i) =\sum_{i=1}^r \rho(U_i)$ and 
 $\rho^*(\bigcup_{i=1}^r U_i)=\sum_{i=1}^r \rho^*(U_i)$ 
\end{corollary}

{\bf Proof.}
Let $E' \subseteq E(H)$  be a set of hyperedges covering $U_1 \cup \dots \cup U_r$
with $|E'|=\rho(U_1 \cup \dots \cup U_r)$.
For each $u \in U_1 \cup \dots \cup U_r$, let $E'_u=\{e'  \in E'| u \in e'\}$.
For each $1 \leq i \leq r$, let $E'_i=\bigcup_{u \in U_i} E'_u$.
Due to the minimality of $E'$, $E'_1 \cup \dots \cup E'_r=E'$.
According to Proposition \ref{decompsep},
$E'_i \cap E'_j=\emptyset$ for each $1 \leq i \neq j \leq r$. 
Thus $|E'|=|E'_1|+ \dots +|E'_r|$. 
Observe that each $E'_i$ is the smallest set of edges covering $U_i$
Indeed, if there $E''_i$ covering $U_i$ such that $E''_i$ is smaller than $E'_i$
then, assuming w..o.g. that $i=1$, we observe that
$E''_1 \cup E'_2+ \dots+E'_r$ covers $U_1 \cup \dots \cup U_r$ and that
$|E''_1 \cup E'_2 \cup \dots \cup E'_r| \leq |E''_1|+|E'_2|+ \dots+|E'_r|<|E'|=\rho(U_1 \cup \dots \cup U_r)$
We thus conclude that $|E'_i|=\rho(U_i)$ for each $1 \leq i \leq r$ and hence
$\rho(U_1 \cup \dots \cup U_r)=\rho(U_1)+ \dots+\rho(U_r)$. 

For the fractional case, let $\gamma:E \rightarrow [0,1]$ be a witnessing function. 
Let $E'$ be the set of all hyperedges whose weight is positive.
With the notation as in the previous paragraph we observe that, due to the minimality
of $\gamma$, $E'=E'_1 \cup \dots \cup E'_r$. 
Arguing as in the previous paragraph, we observe that the union is disjoint.
Therefore, $\rho^*(E')=\sum_{e \in E'} \gamma(e)=\sum_{e \in E'_1} \gamma(e)+ \dots+\sum_{e \in E'_r} \gamma(e)$.
Arguing as in the previous paragraph, we observe that 
$\sum_{e \in E'_i} \gamma(e)=\rho^*(U_i)$ for each $1 \leq i \leq r$.
$\blacksquare$

The notion of tree decomposition is closely related to the notion of treewidth.
This notion can be naturally generalized to $f$-width where $f$ is an arbitrary
function from $2^{V(H)}$ to reals. 
Under this definition, for a tree decomposition $(T,{\bf B})$ of a hypergraph $H$,
the $f$-width of a bag ${\bf B}(x)$ is simply $f({\bf B}(x))$. The
$f$-width of $(T,{\bf B})$ is $max_{x \in V(T)} f({\bf B}(x))$.
Finally, the $f$-width of $H$ is the smallest $f$-wdith of a tree decomposition of $H$. 
In this framework, the treewidth is the $f$-width where $f(S)=|S|-1$. 
The $\rho$-width is called \emph{generalized hypertree width} (\textsc{ghw})
and the $\rho^*$-width is called \emph{fractional hypertree wdith}  (\textsc{fhw}). 
The respective measures of a graph $H$ are denoted by $ghw(H)$ and $fhw(H)$. 

Courcelle's theorem \cite{CourTheorem}  is an algorithmic meta-theorem stating that
a problem on graphs is \textsc{fpt} parameterized by treewidth if the problem
is definable in Monadic Second Order Logic (\textsc{mso}).
In particular, for a graph $G$ consider a model $(V(G),E(G),U_1, \dots,U_q,v_1,\dots v_r)$.
In this context, $V(G)$ is the universe, $E(G)$ is a binary predicate with $E(u,v)$ being
true if and only if $\{u,v\} \in E(G)$. 
Also, $U_1, \dots, U_q$ are subsets of $V(H)$ and $v_1, \dots, v_r$ are elements of $V(H)$.
An \textsc{mso} formula can be seen as an upgrade of a first order formula where 
quantification is allowed over subsets of the universe. 
Also, inside a formula, such a subset $U$ is treated as a unary predicate where
$U(v)$  is true if  and only $v \in U$.
We also consider $U_1 \subseteq U_2$ as a 'built-in' predicate as it can be easily defined
as $\forall v~U_1(v) \rightarrow U_2(v)$.

The Courcelle's theorem states that for any \textsc{mso} sentence $\varphi$ there is a function
$f_{\varphi}$ so that $(V(G),E(G),U_1, \dots,U_q,v_1,\dots v_r) \models \varphi$
can be tested in $O(f_{\varphi}(k) \cdot (|V(G)|+|E(G)|))$ where $k$ is an upper bound 
on the treewidth of $G$. Note that the runtime depends on $\varphi$, so its size must
itself be upper bounded by a function of $k$.

\section{Computation of generalized hypertree width}
This section consists of four subsections. 
In the first subsection we present a generic algorithm for constant approximation of \textsc{ghw}
based on computation of balanced separators. This algorithm is an adaptation
of the treewidth approximation approach as presented in \cite{GraphMinors13} and \cite{ReedTWD}.
The proposed algorithm computes \textsc{ghw} of the input graph $H$ within
a polynomial number of its recursive applications. However, each \emph{individual}
recursive application involves a polynomial number of runs of two auxiliary functions:
one testing existence of a balanced separator 
and one testing whether the edge cover number of the given set is at most the parameter $k$. 
Thus the proposed algorithm is a generic framework for constant \textsc{ghw} approximation
and its runtime for a specific class of hypergraphs depends on the runtime of the auxiliary functions
for this class. 

In the remaining subsections, we show that for hypergraphs of rank at most $r$,
the auxiliary functions are \textsc{fpt} parameterized by $k$ and $r$. 
As the edge cover testing can be done in a brute-force manner, we concentrate on computation
of balanced separators. This is done in the second subsection. The third subsection proves an auxiliary theorem
needed for the second subsection. In the last subsection we prove the resulting theorem 
that \textsc{ghw} of graphs of bounded rank admits constant ratio \textsc{fpt} approximation parameterized by $k$ and
$r$. 

\subsection{The framework}
The cornerstone of the proposed approach is a theorem stating that
a small \textsc{ghw} implies existence of balanced separators with a small edge cover number.
Definition \ref{baldef} and Theorem \ref{balancedsep} provided below are adaptations of similar 
definitions and statements in
the context of treewidth computation.

\begin{definition} \label{baldef}
Let $H$ be a hypergraph and $X,S \subseteq V(H)$.
We say that $X$ is a \emph{balanced separator} for $S$
if for each connected component $V'$ of $H \setminus X$,
$\rho(V' \cap S)<2/3 \rho(S)$.
\end{definition}

\begin{theorem} \label{balancedsep}
Suppose that $ghw(H) \leq k$. 
Let $S \subseteq V(H)$ with $\rho(S) \geq 3k+1$.
Then $S$ has a balanced separator $X$ such that $\rho(X) \leq k$.  
\end{theorem} 

{\bf Proof.}
Let $(T,{\bf B})$ be a tree decomposition of $H$ of \textsc{ghw} at most $k$.
We are going to demonstrate that there is $x \in V(T)$
so that for each connected component $V'$ of $T \setminus x$,
$\rho(({\bf B}(V') \setminus {\bf B}(x)) \cap S) <2/3 \rho(S)$.
The statement will immediately follow from Proposition \ref{decompsep}. 
Pick an arbitrary node $x \in V(T)$ and run the following algorithm.

\begin{algorithm}
\caption{$FBC(H,T,{\bf B},x)$ (the name stands for 'Find Balanced Separator')}
\begin{algorithmic}
\State $i \leftarrow 1$
\State $x_i \leftarrow x$
\State $T_i \leftarrow T$
\While {$T_i \setminus x_i$ contains a component $T'$
          with $\rho(({\bf B}(T') \setminus {\bf B}(x_i)) \cap S) \geq 2/3 \rho(S)$}
  \State Let $x'$ be the neighbour of $x_i$ in $T'$
  \State $i \leftarrow i+1$
  \State $x_i \leftarrow x'$
  \State $T_i \leftarrow T'$
\EndWhile
\end{algorithmic}
\end{algorithm}

\begin{claim}  \label{clm1}
The above algorithm stops.
Moreover, let $q$ be the last value of $i$. 
explored by the algorithm.  Then  $T_q$ contains more than one node.
\end{claim}

{\bf Proof.}
Note first that $T$ contains more than one node.
Indeed, otherwise, by definition of \textsc{ghw}, $\rho({\bf B}(T)) \leq k$.
As $S \subseteq V={\bf B}(T)$, we get a contradiction to the definition of $S$.
(In fact, $T$ must contain at least $4$ nodes.)

If the algorithm stops at $i=1$ then $T_i=T$ and we are done. 
Otherwise, by construction, for each $i>1$,  $|V(T_i)|<|V(T_{i-1})|$. 
So, if the algorithm does not stop, it eventually gets into a situation where $|V(T_i)|=1$
and $V(T_i)$ contains a subset of $S$ with the edge cover number greater than $2k$.
But this is a contradiction due to the same argument as in the first paragraph. 
So, the algorithm must stop earlier than that.
$\square$

\begin{claim} \label{clm2}
For each $i$ considered by the algorithm, there is at most one component
$T'$ of $T_i \setminus x_i$ with $\rho(({\bf B}(T') \setminus {\bf B}(x_i)) \cap S) \geq 2/3\rho(S)$.
Moreover, let $T'_1, \dots, T'_r$ be the connected components of $T_i \setminus x_i$
and assume w.l.o.g. that $\rho(({\bf B}(T'_1) \setminus {\bf B}(x_i)) \cap S) \geq 2/3 \rho(S)$. 
Then $\rho((({\bf B}(T'_2) \cup \dots \cup {\bf B}(T'_r)) \setminus {\bf B}(x_i))\cap S) \leq 1/3 \rho(S)$.   
\end{claim}

{\bf Proof.}
Observe that $\rho(S) \geq \rho(S \cap ({\bf B}(T'_1) \cup \dots \cup {\bf B}(T'_r) \setminus {\bf B}(x)))=
\rho(S \cap {\bf B}(T'_1) \setminus {\bf B}(x) )+ \dots+ \rho(S \cap {\bf B}(T'_r) \setminus {\bf B}(x))$,
the equality follows from Corollary \ref{decomprho}.  
Then 
$\rho(S \cap ({\bf B}(T'_2) \cup \dots {\bf B}(T'_r) \setminus {\bf B}(x)))=
\rho(S \cap {\bf B}(T'_2) \setminus {\bf B}(x))+ \dots \rho(S \cap {\bf B}(T'_r) \setminus {\bf B}(x)) 
\leq \rho(S)-\rho(S \cap {\bf B}(T'_1) \setminus {\bf B}(x)) \leq 1/3 \rho(S)$.
$\square$

Let $q$ be as specified in Claim \ref{clm1}.
We claim that ${\bf B}(x_q)$ is the desired set $X$.  
If $q=1$, we are done by definition. 
Otherwise, let $T'_1, \dots, T'_r$ be the components of
$T_{q-1} \setminus x_{q-1}$. 
Assume w.l.o.g. that $T_q=T'_1$.
What are the components of $H \setminus {\bf B}(x_q)$?.
By Proposition \ref{decompsep}, some of them are subsets of ${\bf B}(T'') \setminus {\bf B}(x_q)$ where $T''$ is a component of 
$T_q \setminus x_q$.
By the algorithm and the definition of $q$ as the first value for which
the while-condition becomes false, we conclude that 
$\rho(S \cap {\bf B}(T'') \setminus {\bf B}(x_q))<2/3 \rho(S)$. 

The remaining components are subsets of $V^*={\bf B}(x_{q-1}) \cup {\bf B}(T'_2) \cup \dots \cup {\bf B}(T'_r) \setminus {\bf B}(x_q)$.
Harnessing Claim \ref{clm2}, we conclude that
$\rho(S \cap V^*) \leq \rho (S \cap {\bf B}(x_{q-1}))+\rho(S \cap ({\bf B}(T'_2) \cup \dots \cup {\bf B}(T'_r) \setminus {\bf B}(x_q))) \leq k+1/3 \rho(S)$.
By definition of $S$, $k<1/3 \rho(S)$, hence $\rho(S \cap V^*)<2/3 \rho(S)$.
$\blacksquare$


We now define 
an algorithm $ApproxGHW(H,S,k,p)$, where $H$ is a hypergraph and $S \subseteq V(H)$. 
The algorithm returns either a tree decomposition $(T,{\bf B})$ with a bag containing $S$ or 'NO'.
In the latter case, it is guaranteed that $ghw(H)>k$.

We need two auxiliary functions.
The first, $BalSep(H,S,k)$, returns a balanced separator $X$ of $S$ with $\rho(X) \leq p$. 
or 'NO' if such a separator does not exist. Note that if $\rho(S) \geq 3k+1$ then 'NO'
implies that $ghw(H)>k$ by Theorem \ref{balancedsep}. 
The second function is $IsCover(H,S,p)$. The function returns $True$ if $S$ has an edge cover (in $H$)
of size at most $p$ and $False$ otherwise. 
 
The algorithms for the functions and their respective runtimes 
will depend on classes for graphs for which we run $ApproxGHW$.
In this section, we provide a generic description treating these functions as oracles.
\begin{algorithm}
\caption{$ApproxGHW(H,S,k,p)$: returns a tree decomposition $(T,{\bf B})$ with $S$ being a subset of a bag or 'NO'}
\begin{algorithmic}
\If {$IsCover(H,V(H),p)$ is $True$}  
  \State Return $(T, {\bf B})$ where $T$ consists of a single node $x$ and ${\bf B}(x)=V(H)$.
\EndIf
\While {$IsCover(H,S,p)$ is $True$} 
  \State Let $S \leftarrow S \cup \{v\}$ where $v$ is an arbitrary element of $V(H) \setminus S$.
\EndWhile
\State $X \leftarrow BalSep(H,S,k)$.
\State If $X$ is 'NO' then return 'NO'.
\State Let $V_1, \dots, V_q$  be the connected components 
of $H \setminus X$  such that $\rho(V_i) \geq 3k$.  
\For {each $1 \leq i \leq q$}
  \State $H_i \leftarrow H[V_i \cup X]$ for each $1 \leq i \leq q$.
  \State $O_i \leftarrow ApproxGHW(H_i,(S \cap V_i) \cup X,k,p)$.
  \State If $O_i$ is 'NO' then return 'NO' (otherwise
 $O_i=(T_i,{\bf B_i})$, a tree decomposition of $H_i$ with a node $x_i$ 
such that $(S \cap V_i) \cup X \subseteq {\bf B_i}(x_i)$).
\EndFor
\State Introduce a new node $x$ and make it adjacent to $x_1, \dots, x_q$.
Let $T'$ be the resulting tree. 
\State For each connected component $V'$ of $H \setminus X$ with $\rho(V')<3k$,
introduce a new node $x'$ and connect it by edge to $x$. Let $T$ be the resulting tree.  
\State Let ${\bf B}: V(T) \rightarrow 2^H$ be defined as follows.
   \begin{itemize}
   \item ${\bf B}(x)=S \cup X$. 
   \item For each $1 \leq i \leq q$ and for each $y \in V(T_i)$,
              ${\bf B}(y)={\bf B_i}(y)$.
  \item For each $x'$ corresponding to a component $V'$ of $H \setminus X$ with $\rho(V')<3k$,
             ${\bf B}(x')=V' \cup X$.
 \end{itemize}
\State Return $(T,{\bf B})$. 
\end{algorithmic}
\end{algorithm}

\begin{lemma} \label{lem111}
Let $X$ be a balanced separator for $S$ with $\rho(X) \leq k$.
Then the following statements hold.
\begin{enumerate}
\item 
Suppose that $\rho(S) \geq 3k+1$.
Then $H \setminus X$ has at least two connected components. 
\item 
Assume, in addition, that $\rho(S) \geq 6k+1$.
Let $V'$ be a connected component of $H \setminus X$.
Then $\rho(V' \cup X)<\rho(V(H))$.
\end{enumerate} 
\end{lemma}

{\bf Proof.}
For the first item, assume that $H \setminus X$ is connected.
Then $\rho(S) \leq \rho((V(H) \setminus X) \cap S)+\rho(X \cap S) \leq \rho((V(H) \setminus X) \cap S)+k$.
That is $\rho(S)-k \leq \rho((V(H) \setminus X) \cap S)$. It follows from the definition of $S$ that
$\rho((V(H) \setminus X) \cap S) \geq 2/3 \rho(S)$, a contradiction.

Let us now prove the second statement. 
Let $V''=V(H \setminus X) \setminus V'$. By the first statement $V''$ is non-empty.
By Corollary \ref{decomprho},
$\rho(V' \cup V'')=\rho(V')+\rho(V'')$. 
Assume that $\rho(V' \cup X)=\rho(V(H))$,
Clearly $\rho(V' \cup X) \leq \rho(V')+\rho(X) \leq \rho(V')+k$.
It follows that $\rho(V(H))-\rho(V') \leq k$. 
On the other hand, $\rho(V(H)) \geq \rho(V' \cup V'')$ and
hence $\rho(V(H))-\rho(V') \geq \rho(V'')$. We conclude that $\rho(V'') \leq k$.

Now, $\rho(S) \leq \rho(S \cap V')+ \rho(S \cap X)+ \rho(S \cap V'') \leq \rho(S \cap V')+\rho(X)+\rho(V'') \leq \rho(S \cap V')+2k< 2/3\rho(S)+2k$.
We conclude that $2k>1/3\rho(S)$ in contradiction to our assumption about $S$.
$\blacksquare$

\begin{lemma} \label{lem112}
Let $k,a_1, \dots, a_q$ be integers such that $k \geq 1$, $q \geq 2$, and
$a_i \geq 3k$ for each $1 \leq i \leq q$.  
Then $\sum_{i=1}^q (a_i+k)^2<(\sum_{i=1}^q a_i)^2$
\end{lemma}

{\bf Proof.}
Let $r_i=a_i/k$ for $1 \leq i \leq q$. 
Assume first that $q=2$. 
Then we need to show that $(r_1k+k)^2+(r_2k+k)^2 < (r_1k+r_2k)^2$. 
On the left-hand side, we obtain
$r_1^2k^2+r_2^2k^2+2r_1k^2+2r_2k^2+2k^2$.
On the right-hand side we obtain 
$r_1^2k^2+r_2^2k^2+2r_1r_2k^2$.
After removal of identical items and dividing both parts
by $2k^2$, it turns out that we need to show that
$r_1+r_2+1<r_1r_2$. But it is not hard to see that that this is indeed so
if both $r_1$ and $r_2$ are at least $3$.

Assume now that $q>2$. 
Then by the previous paragraph,
$\sum_{i=1}^{q-2} (a_i+k)^2+(a_{q-1}+k)^2+(a_q+k)^2<\sum_{i=1}^{q-2} (a_i+k)^2+(a_{q-1}+a_q)^2<
\sum_{i=1}^{q-2} (a_i+k)^2+(a_{q-1}+a_q+k)^2<(\sum_{i=1}^q a_i)^2$, the last inequality follows
from the induction assumption.
$\blacksquare$

\begin{lemma} \label{lem113}
Let $k \geq 1$ and suppose that $H$ is connected hypergraph with $\rho(H) \geq 1$
Let $S \subseteq V(H)$.
Then $ApproxGHW(H,S,k,6k)$ invokes the function $ApproxGHW$ at most $\rho(V(H))^2$ 
times. 
\end{lemma}

{\bf Proof.}
By induction on $\rho(H)$.
If $\rho(H) \leq 6k$ then there is only one invocation so we are done. 
Otherwise, for the function to run more than once, there need to be
components $V_1, \dots V_q$ with $\rho(V_i) \geq 3k$ for $1 \leq i \leq q$.
Let $H_i=H[V_i \cup X]$ for $1 \leq i \leq q$. 
By Lemma \ref{lem112}, $\rho(V_i \cup X) <\rho(V(H))$.
Taking into account that $\rho_{H_i}(V_i \cup X)=\rho_H(V_i \cup X)$ and applying the induction
assumption, we conclude that the recursive call
$ApproxGHW(H_i,V_i \cup X,k,6k)$ invokes at most $\rho(V_i \cup X)^2$ 
applications of $ApproxGHW$. We thus need to demonstrate
that $\sum_{i=1}^q \rho(V_i \cup X)^2+1 \leq \rho(V(H))^2$ (the final $1$ 
accounts for the starting application $ApproxGHW(H,S,k,6k)$).
We prove an equivalent statement that 
$\sum_{i=1}^q \rho(V_i \cup X)^2 < \rho(V(H))^2$. 

Assume first that $q \geq 2$. 
Taking into account that for each $1 \leq i \leq q$,
$\rho(V_i \cup X) \leq \rho(V_i)+\rho(X) \leq \rho(V_i)+k$, it is sufficient to prove 
$\sum_{i=1}^q (\rho(V_i)+k)^2 < \rho(V(H))^2$. 
By Corollary \ref{decomprho},
$\sum_{i=1}^q \rho(V_i)=\rho(\bigcup_i^q V_i) \leq \rho(V(H))$. 
That is, it is sufficient to prove that
$\sum_{i=1}^q (\rho(V_i)+k)^2 < (\sum_{i=1}^q \rho(V_i))^2$.
But this is immediate from Lemma \ref{lem112}.
It remains to assume that $q=1$.
Then $\rho(V_i \cup X)^2<\rho(V(H))^2$.
is immediate from the second item of Lemma \ref{lem111}.
$\blacksquare$

\begin{lemma} \label{lem114}
Let $k \geq 1$ and suppose that $H$ is connected hypergraph with $\rho(H) \geq 1$
Let $S \subseteq V(H)$.
Then $ApproxGHW(H,S,k,3k)$ invokes the function $ApproxGHW$ at most $|V(H)| \cdot \rho(V(H))^2$ 
times. 
\end{lemma}

{\bf Proof.}
The recursive calls of $ApproxGHW$ function can be naturally organized into a rooted tree $R$.
The initial application corresponds to the root of the tree. The children of the given 
application correspond to the recursive calls made directly within this application.
The applications that do not invoke any new recursive applications correspond to the 
leaves of the tree. 

\begin{claim}
$R$ has at most $\rho(V(H))^2$ leaves.
\end{claim}

{\bf Proof.}
By induction on $|V(H)|$.
If $|V(H)|=1$ then the function does not call itself recursively.
If the considered application has only one child
then, by the first item of Lemma \ref{lem111},
the graph in the recursive application from the child has a smaller number
of vertices. The statement holds for the child application by the induction
assumption, hence it holds for the parent application since the graph there is
larger and the sets of leaves are the same. 

If the considered application has two or more children
then we apply the same reasoning as in the last paragraph
of Lemma \ref{lem113} for the case of $q \geq 2$.
$\square$

It remains to observe that the length of any 
root-leaf path is at most $V(H)$. This is because the number of vertices of the input hypergraph of a child
is always smaller than the number of vertices of the input hypergraph of the parent according
to the first item of Lemma \ref{lem111}.
$\blacksquare$

\begin{theorem} \label{ghwpattern}
Let ${\bf H}$ be a class of hypergraphs.
Suppose that for each $H \in {\bf H}$, 
a single recursive application of $ApproxGHW$ takes $x(H,k)$.
Let $H \in {\bf H}$ be a connected hypergraph. 
\begin{enumerate}
\item $ApproxGHW(H,\emptyset,k,3k)$ returns a tree decomposition 
of \textsc{ghw} at most $4k$ or 'NO'. In the latter case, it is guaranteed that
$ghw(H)>k$. The runtime of the function is $O(x(H,k) \cdot |V(H)| \cdot \rho(H)^2)$.
\item  $ApproxGHW(H,\emptyset,k,6k)$ returns a tree decomposition 
of \textsc{ghw} at most $6k$ or 'NO'. In the latter case, it is guaranteed that
$ghw(H)>k$. The runtime of the function is $O(x(H,k) \cdot \rho(H)^2)$.
\end{enumerate}
\end{theorem}

{\bf Proof.}
The runtime for the first item follows from Lemma \ref{lem114}
and the runtime for the second item follows from Lemma \ref{lem113}. 

Next, it is easy to verify by induction that if the algorithm returns NO
then there is $S \subseteq V(H)$ with $\rho(H)>3k$ that does not have a balanced
separator of size at most $k$.  It follows from Theorem \ref{balancedsep}, that in this 
case, $ghw(H)>k$.

Let us now verify that if $ApproxGHW(H,S,k,p)$ does not return 'NO' then
it returns a tree decomposition $(T,{\bf B})$ of $H$
and such one that there is a node $x \in V(T)$ with $S \subseteq {\bf B}(x)$.

For a single node this follows by construction, so we assume existence of a set $X$
as specified by the algorithm.
Let $V_1, \dots, V_q$ be the components of $H \setminus X$ where $\rho(V_i) \geq 3k$
and let $V'_1, \dots V'_r$ be the components of $H \setminus X$ where $\rho(V_i)<3k$.
Let $S^*$ be the superset of $S$ obtained at the end of the first while-loop. 
For each $1 \leq i \leq q$, the algorithm recursively applies
$ApproxGHW(H_i=H[V_i \cup X],(S^* \cap V_i) \cup X,k,p)$ and returns a pair 
$(T_i,{\bf B_i})$. By the induction assumption, $(T_i, {\bf B_i})$ is a tree decomposition
of $H[V_i \cup X]$ so that there is $x_i \in T_i$ with $(S^* \cap V_i) \cup X \subseteq {\bf B_i}(x_i)$.
The algorithm then constructs $T$ by introducing a new node $x$, connecting it to each $x_i$
and then connecting it to new nodes $x'_1, \dots x'_r$ corresponding to the components 
$V'_1, \dots, V'_r$. Clearly $T$ is a tree. By construction, $S \subseteq S^* \subseteq {\bf B}(x)$.
It remains to verify that ${\bf B}$ satisfies the axioms of tree decomposition.

Let $u \in V(H)$. Then there is $1 \leq i \leq q$ such that $u \in V_i \cup X$ 
or there is $1 \leq i \leq r$ such that $u \in V'_i \cup X$.
In the former case, by the induction assumption, there is $x^* \in V(T_i)$ such
that $u \in {\bf B_i}(x^*)={\bf B}(x^*)$. In the latter case $u \in {\bf B}(x_i)$ simply
by construction. Thus we have verified the union axiom.
Let $e \in E(H)$. Then, by Proposition \ref{decompsep}, either $e \in E(H_i)$ for some $1 \leq i \leq q$
or $e \subseteq V'_i \cup X$ for some $1 \leq i \leq r$. Thus the containement axiom
is verified by the same argument.

Let $u \in X$. By the induction assumption, for each $1 \leq i \leq q$, the nodes of $T_i$ whose bags in ${\bf B_i}$
contain $u$ induce a connected subgraph $T^*_i$ of $T_i$. As the projection
of ${\bf B}$ to $V(T_i)$ is ${\bf B_i}$, the same is true regarding ${\bf B}$.
Moreover, $x_i \in V(T^*_i)$. Furthermore, by construction $u \in {\bf B}(x)$
and $u \in {\bf B}(x'_i)$ for all $1 \leq i \leq r$. As $x$ is connected to $x_1,\dots x_q$
and to $x'_1, \dots x'_r$, we conclude that the nodes of $T$ whose bags in ${\bf B}$
contain $u$ induce a connected subgraph of $T$. 

Let $u \in V(H) \setminus X$. Then there is at precisely one set among $V_1, \dots, V_q, V'_1, \dots, V'_r$
that contains $u$. If the set is some $V_i$ then we let $T^*_i$ to be as in the previous paragraph
and the set of nodes of $T$ with bags containing $u$ is either $V(T^*_i)$ or $V(T^*_i) \cup \{x\}$,
inducing a connected subgraph of $T$  in both cases (note that, in the latter case, $u \in {\bf B}(x_i)$). 
If the set is some $V'_i$ then $u$ is contained in the bags of $x$ and $x'_i$ or in the bag of $x'_i$ only
and $x$ is adjacent to $x'_i$ by construction. So, the connectivity property has been established.

It remains to prove the required upper bound on the edge cover numbers
of the bags in the considered tree decomposition.
Let us prove first that $ApproxGHW(H,S,k,3k)$ returns a tree decomposition of \textsc{ghw} at most $4k$
as long as $\rho(S) \leq 3k$ (this will clearly imply the first statement of the theorem). 
If $|V(H)|=1$, or, more generally if $\rho(V(H)) \leq 3k$, this follows by construction.
Otherwise, note that 
$\rho(S^*)=3k+1$. In particular, it follows from the definition of $X$ that for each connected
component $V'$ of $H \setminus X$, $\rho(V' \cap S^*) \leq 2k$. 
In particular, this means that if $V'$ is one of $V_1, \dots V_q$ as above then
the second parameter of the recursive application $ApproxGHW(H[V_i \cup X], (S^* \cap V_i) \cup X,k,3k)$,
is at most $3k$.
By the induction assumption, the \textsc{ghw} of $(T_i,{\bf B_i})$ is at most $4k$.
It remains to observe that, by construction, the edge cover numbers ${\bf B}(x)$
and of nodes ${\bf B}(x'_i)$ corresponding to components $V'_1, \dots V'_r$ as above
are all at most $4k$.  

For the second statement, we consider $ApproxGHW(H,S,k,6k)$ with $\rho(S) \leq 5k$.
The reasoning is similar to the previous paragraph. The essential (numerical) difference is that
the first loop produces an $S^*$ with $\rho(S^*)=6k+1$ implying
that for each $1 \leq i \leq q$, $\rho(V_i \cap S^*) \leq 4k$ and implying, in turn that the edge
cover number of the second parameter of any recursive application is at most $5k$.
$\blacksquare$

\subsection{Balanced separator computation}
In this section we demonstrate that testing existence of a balanced separator 
with edge cover number at most $k$ for a set $S$ with $\rho(S)=O(k)$ is \textsc{fpt} parameterized by $k$ and the rank $r$ 
of the input hypergraph $H$. The main engine of the resulting algorithm is an auxiliary \textsc{fpt} procedure
(in $k$ and $r$) testing existence of a $s,t$-cut of edge cover at most $k$.
The algorithm for this procedure is described in the next subsection.

To test existence of a balanced separator, we utuilize the fact that $|S|=O(kr)$
and consider, in a brute force manner, all non-empty disjoint subsets
$S_1$ and $S_2$ of $S$ with $\rho(S_i)<2/3\rho S$ for each $i \in \{1,2\}$.
For each such a pair $S_1,S_2$, we use the auxiliary procedure to test existence of an $S_1,S_2$-cut $X$
with $X \cap S=S \setminus (S_1 \cup S_2)$ and with $\rho(X) \leq k$.
A detailed description is provided below.

\begin{lemma} \label{lemmul3}
Let $H$ be a hypergraph, $S \subseteq V(H)$ so that 
$\rho(S)>3k$ is not a multiple of $3$. 
Let $X$ be a balanced separator for $S$ with $\rho(X) \leq k$.
Then $H \setminus X$  can be seen as the union
of two vertex-disjoint subgraphs $H_1$ and $H_2$ (note that since
$H_1 \cup H_2=H \setminus X$, these subgraphs are also edge-disjoint)
so that $\rho(S \cap V(H_i))<2/3 \rho(S)$ for each $i \in \{1,2\}$.
\end{lemma}

{\bf Proof.}
Let $H'_1, \dots H'_q$ be the components of $H \setminus X$.
By Lemma \ref{lem111}, $q>1$.  
If $q=2$, we are done by definition of a balanced separator. 
If, say $\rho(S \cap V(H'_1)) > 1/3 \rho(S)$ then
let $H_1=H'_1$ and $H_2=H'_2 \cup \dots \cup H'_q$.
It remains to assume that for each $1 \leq i \leq q$
$\rho(S \cap V(H'_i)<1/3 \rho(S)$ (because $\rho$ is an integer function and hence 
by assumption about $\rho(S)$, the value of a $\rho$ function cannot be equal
$\rho(S)/3$ and that $q>2$. 
Then apply this argument inductively to 
$H'_1 \cup H'_2,H'_3, \dots, H'_q$. 
$\blacksquare$

\begin{definition}
Let $H$ be a hypergraph, $\emptyset \subset U \subseteq V(H)$.
We define a \emph{contraction} of $U$ in $H$ a hypergraph,
where the set $U$ is replaced by a single vertex $u$.
Formally,
$V(H_{U \leftarrow u})=(V(H) \setminus U) \cup \{u\}$
and
$E(H_{U \leftarrow u})=\{g(e)|e \in E(H)\}$
where $g(e)$ is defined as follows.
If $e \cap U \neq \emptyset$ then 
$g(e)=(e \setminus U)  \cup \{u\}$.
Otherwise, $g(e)=e$. 
\end{definition}

In case of two disjoint sets $U_1,U_2 \subseteq V(H)$,
we write $H_{U_1 \leftarrow u_1,U_2 \leftarrow u_2}$
instead of $[H_{U_1 \leftarrow u_1}]_{U_2 \leftarrow u_2}$. 

\begin{definition}
Let $S \subseteq V(H)$. 
Let $S_1,S_2$ be two non-empty disjoint subsets of $S$.
We denote by $H^S_{S_1 \leftarrow s_1.S_2 \leftarrow s_2}$
the graph with the set of vertices 
$V(H_{S_1 \leftarrow s_1, S_2 \leftarrow s_2})$ and
the set of edges $E(H_{S_1 \leftarrow s_1,S_2 \leftarrow s_2}) \cup 
\{\{s_1,s\}, \{s_2,s\}|s \in S \setminus (S_1 \cup S_2)\}$. 
\end{definition}

\begin{lemma} \label{lem22}
Let $H$ be a hypergraph, $S \subseteq V(H)$, $S_1$ and $S_2$ be two non-empty disjoint subsets of $S$.
Denote $H^S_{S_1 \leftarrow s_1,S_2 \leftarrow s_2}$ by $H^*$.
Let $X \subseteq V(H^*) \setminus \{s_1,s_2\}$ 
Then the following two statements hold.
\begin{enumerate}
\item $X$ is an $s_1,s_2$-separator of $H^*$ if and only if $X \cap S=S \setminus (S_1 \cup S_2)$
and $X$ is an $S_1,S_2$-cut of $H$.
\item $\rho_H(X)=\rho_{H^*}(X)$. 
\end{enumerate}

\end{lemma}

{\bf Proof.}
Let $X \subseteq V(H)$ be 
an $S_1,S_2$-cut with  $X \cap S=S \setminus (S_1 \cup S_2)$.
Then $X$ separates $s_1$ and $s_2$ in $H^*$.
Indeed, suppose that there is a path $P$ from $s_1$ to $s_2$ in $H^* \setminus X$.
Due to $S \setminus (S_1 \cup S_2) \subseteq X$, $P$ is also a path in $H_{S_1 \leftarrow s_1,S_2 \leftarrow s_2}$. 
Let $P=v_1,e_1, \dots, v_{a-1},e_{a_1},v_a$ where $s_1=v_1, \dots v_a=s_2$ is a sequence of
vertices and $e_1, \dots,e_{a-1}$ are hyperdges of $H^*$ so that 
for each $1 \leq i \leq a-1$, $v_i,v_{i+1} \in e_i$.
By definition of $H_{S_1 \leftarrow s_1,S_2 \leftarrow s_2}$,
replacing $e_1, \dots e_{a-1}$ with their respective preimages, 
we transform $P$ into a path from some $s'_1 \in S_1$ to some $s'_2 \in S_2$
in $H$ with the same intermediate vertices. Consequently the resulting path does not contain
$X$ and hence is a path in $H \setminus X$, a contradiction.

Conversely, assume that $X$ is an $s_1,s_2$-separator of $H^*$.
We note that $S \setminus (S_1 \cup S_2) \subseteq X$ simply by construction.
We also observe that in $H \setminus X$, there is no path from $S_1$ to $S_2$
because the image of the edges of such a path would result in an 
$s_1,s_2$-path of $H^*$ avoiding $X$. Thus we have proved the first statement. 

For the second statement, let $e_1, \dots e_q$ be hyperedges of $H$
covering $X$. By construction, their images in $H^*$ also cover $X$.
Therefore, $\rho_{H^*}(X) \leq \rho_H(X)$. 
Conversely, let $e_1, \dots, e_b$ be a set of hyperedges of $H^*$ covering $X$.
Replace them with hyperedges $e'_1, \dots, e'_b$ of $H$ (not necessarily distinct) as follows.
If $e_i \in E(H^*) \setminus E(H_{S_1 \leftarrow s_1,S_2 \leftarrow s_2})$
then we can identify precisely one element $s(e_i) \in S \cap X'$ covered
by $e_i$. As $H$, by assumption, does not have isolated vertices, 
simply identify an arbitrary hyperedge $e'_i$ containing $s(e_i)$.
Otherwise, let $e'_i$ be an arbitrary hyperedge of $H$ whose image in
$H_{S_1 \leftarrow s_1,S_2 \leftarrow s_2}$ is $e_i$. 
We observe that $e'_1, \dots, e'_b$ cover $X$.
Indeed, let $u \in X$. If $u=s(e_i)$ for some $1 \leq i \leq b$ then $e'_i$
covers $u$ simply by definition. Otherwise, as edges for which $s(e_i)$ are
defined do not cover any elements but $s(e_i)$. $u$ is covered by some 
$e_i \in H_{S_1 \leftarrow s_1, S_2 \leftarrow s_2}$. But then as $u \notin S_1 \cup S_2$,
$u \in e'_i$ simply by construction. Thus we conclude that $\rho_{H}(X) \leq \rho_{H^*}(X)$. 
$\blacksquare$

\begin{theorem} \label{mainghw21}
Let $H$ be a hypergraph of rank at most $r$. 
Let $s,t$ be two vertices of $H$.
Then testing existence of an $s,t$-separator $X$ with $\rho(X) \leq k$
is \textsc{fpt} parameterized by $k$ and $r$.
\end{theorem}

\begin{theorem} \label{mainghw22}
Let $H$ be a hypergraph of rank at most $r$.
Let $k$ be a positive integer. 
Let $S \subseteq V(H)$ with $\rho(S)=3k+1$ or $\rho(S)=6k+1$.
Then testing existence of a balanced separator for $S$ 
is \textsc{fpt}  parameterized by $k$ and $r$.
\end{theorem}'

{\bf Proof.}
Consider all possible non-empty disjoint subsets $S_1$ and $S_2$ 
of $S$ with $\rho(S_i)<2/3 \rho(S)$ for each $i \in \{1,2\}$.
Since $|S|=O(kr)$, all possible sets can be considered in a brute-force
manner and, for the given pair of sets, their $\rho$ numbers can also
be considered in a brute-force manner applied to $H[S_1]$
and to $H[S_2]$.

For each pair $S_1$ and $S_2$ of sets as above,
compute $H^*=H^S_{S_1 \leftarrow s_1,S_2 \leftarrow s_2}$
and test the existence of an $s_1,s_2$-separator $X$ of $H^*$
with $\rho(X) \leq k$. By Theorem \ref{mainghw21},
the testing is \textsc{fpt}.
If such a separator exists for at least one pair $S_1,S_2$ then
return $True$. Otherwise, return $False$.  

It remains to demonstrate that the algorithm returns $True$ if and only
if $S$ has a balanced separator with the edge cover number at most $k$. Let $S_1,S_2$ be the witnessing sets
and let $H^*$ be the corresponding graph for which there exists an $s_1,s_2$-separator
$X$ with $\rho_{H^*}(X) \leq k$. 
By Lemma \ref{lem22}, $S \setminus (S_1 \cup S_2) \subseteq X$ and $\rho_H(X) \leq k$.
We conclude that $X$ is a balanced separator for $H$ with the required upper bound on the edge cover number.
Conversely, suppose that $H$ has a balanced separator $X$ of $S$ with edge cover number at most $k$. 
Our assumption about $\rho(S)$ satisfies the premises of Lemma \ref{lemmul3}.
Let $H_1$ and $H_2$ be the hypergraphs as in the lemma. Let $S_1=V(H_1) \cap S$ and $S_2=V(H_2) \cap S$.
Note that both $S_1$ and $S_2$ are non-empty since otherwise, if say $S_2=\emptyset$,
$\rho(S) \leq \rho(S_1)+\rho(S \cap X)<2/3\rho(S)+1/3 \rho(S)=\rho(S)$, 
a contradiction. It follows that the sets $S_1$ and $S_2$ are considered by the algorithm.
It follows from Lemma \ref{lem22} that the respective graph $H^*$ has an $s_1,s_2$-separator
with edge cover number at most $k$.
Hence these sets witness that the algorithm returns $True$.

$\blacksquare$ 

\subsection{Proof of Theorem \ref{mainghw21}}
\begin{definition}
Let $H$ be a hypergraph, $U \subseteq V(H)$. 
The \emph{torso} of $U$ in $H$ denoted by $Torso(H,U)$
is a graph $G$ with $V(G)=U$ and $E(G)$ is the set of
all $\{u,v\}$ such that $H$ has a path from $u$ to $v$ with
all the intermediate vertices laying outside of $U$. 

The \emph{extended torso} of $U$ in $H$ denoted 
by $Torso^*(H,U)$ is obtained from $Torso(H,U)$ as
follows. Let $e_1, \dots e_r$ be all the hyperedges 
of $H$ having a non-empty intersection with $U$.
Introduce $r$ new vertices $u_1, \dots u_r$.
Make each $u_i$ adjacent to all the vertices of $U \cap e_i$.
\end{definition}

 The following observations will be important for the further reasoning. 

\begin{lemma} \label{smallprop}
\begin{enumerate}
\item Each $u \in V(Torso^*(H,U)) \setminus V(Torso((H,U))$ 
is a simplicial vertex. To put it differently, for every distinct $v_1,v_2$ adjacent 
to $u$ in $Torso^*(H,U)$, $v_1$ and $v_2$ are adjacent in $Torso(H,U)$. 
\item Let $u_1,u_2 \in U$ and $X \subseteq U \setminus \{u_1,u_2\}$.
Then $X$ is a $u_1,u_2$-separator in $Torso(H,U)$ if and only if 
$X$ is a $u_1,u_2$-separator of $Torso^*(H,U)$. 
\item $tw(Torso^*(H,U)) \leq Torso(H,U)+1$.
\item Let $G$ be the incidence graph of $H$. Then $Torso(H,U)=Torso(G,U)$. 
\end{enumerate}
\end{lemma}

{\bf Proof.}
For the first statement, just note that if $v_1$ and $v_2$ belong to the same hyperedge
then $v_1,v_2$ is a path of $H$ without any intermediate vertices. hence the adjacency condition
for a torso is vacuously satisfied. 

The second statement is more convenient to prove in a negative form: that $X$ is not a 
$u_1,u_2$-separator of $Torso^*(U,H)$ if and only if the same is true for $Torso(U,H)$.
Since the latter is a subgraph of the former, we only need to verify that 
any $u_1,u_2$-path $P$ of $Torso^*(H,U)$ avoiding $X$ can be transformed into such a
path of $Torso(H,U)$. Note that the first and the last vertices of $P$ belong to $V(Torso(H,U))$. 
By the first statement, any vertex of $V(Torso^*(H,U)) \setminus V(Torso(H,U))$ occurring on $P$
can be replaced by an edge between its immediate successor and the immediate predecessor.

For the third statement, let $(T,{\bf B})$ be a tree decomposition
of $Torso(H,U)$ of width $tw(Torso(H,U))$. Because of the first statement,
the vertex adjacent to each $u \in V(Torso^*(H,U)) \setminus V(Torso(H,U))$
form a clique and hence there is $x_u \in V(T)$ such that $N(u) \subseteq {\bf B}(x_u)$.
Create a new node $x'_u$, make it adjacent to $x_u$ and let the bag of $x'_u$
be $N[u]$. 

For the fourth statement, let $P=(v_1, \dots, v_r)$ be a path of $H$ between $u_1,u_2 \in U$ 
with all the intermediate vertices being outside $U$.   
Then $G$ has a path $(v_1,e_1, \dots, e_{r-1},v_r)$ where each $e_i$ is the vertex of $G$
corresponding to the hyperedge containing both $v_i$ and $v_{i+1}$. 
Conversely any path between $u_1$ and $u_2$ in $G$ has the form $(v_1=u_1,e_1, \dots, e_{r-1},v_r=u_2)$.
By removal the vertices corresponding to hyperedges we obtain the corresponding path of $H$.
$\blacksquare$

The following theorem is a special case of the Treewidth Reduction Theorem
(Theorem 2.15 of \cite{TALG}). 
\begin{theorem} \label{theorem215}
Let $G=(V,E)$ be a graph, $s,t$ be two distinct vertices and $p \geq 1$
be an integer. Then there is a function $h$ and an \textsc{fpt} algorithm parameterized by $p$
computing a set $C \subseteq V(G)$ 
such that $C$ is a superset of the union of all minimal $s,r$-separators of $G$ of size at most $p$.
Moreover, $tw(Torso(G,C \cup \{s,t\}) \leq h(p)$.
\end{theorem}

Theorem \ref{theorem215} is a tool for design of fixed-parameter algorithms for constrained
separation problem. The rough idea is that if we need to compute a minimal $s,t$-cut subject
to some extra constraints definable in the Monadic Second Order logic (\textsc{mso}), then  the
cut is computed for the torso graph rather than for the original graph. Since the torso graph
has bounded treewidth,  the fixed-parameter tractability of the computation follows from
Courcelle's theorem. At the technical level there are nuances. For example, for some problems,
the torso needs to be modified and it must be proved that the resulting modification does not
significantly increase the treewidth. 

We can harness the Treewidth Reduction Theorem because in hypergraphs of rank at 
most $r$, a separator with edge cover number at most $k$ has size at most $k \cdot r$ .
This means that the treewidth of the respective torso can be upper-bounded by $h(k \cdot r)$. 

\begin{theorem} \label{upgrade}
Let $H$ be a hypergraph of rank at most $r$, $s,t$ be two distinct vertices and $k \geq 1$
be an integer.  
There is a function $g$ and an \textsc{fpt} algorithm parameterized by $k$ and $r$
that returns a set  $C \subseteq V(H)$ such that $C$ is a superset of the
union of all minimal $s,t$-separators of edge cover number at most $k$ and
$tw(Torso(H,C \cup \{s,t\})) \leq g(k,r)$.
\end{theorem}

{\bf Proof.}
Since any subset of $H$ of edge cover number at most $k$ is of 
size at most $k \cdot r$, we can replace the requirement of $C$ containing all
minimal separators of edge cover number at most $k$ by the requirement 
that $C$ should contain all the minimal separators of size at most $k \cdot r$. 
This almost fits the premises of the Treewidth Reduction Theorem 
with the only obstacle that the theorem is not stated for hypergraphs. 
We overcome this obstacle by considering the incidence graph $G$ of $H$
To avoid introducing new notation, the subset of $V(G)$ corresponding to $V(H)$
is also denoted by $V(H)$ and the remaining vertices of $G$ are denoted by $E(H)$. 

\begin{claim}
There is a function $g$ and an \textsc{fpt} algorithm parameterized by $k$ and $r$
that returns a set  $C \subseteq V(G)$ such that $C$ meets the
following constraints.
\begin{enumerate}
\item $C \subseteq V(H)$.
\item $C$ includes all the minimal $s,t$-separators of $G$ of size at most $k \cdot r$ that are subsets of $V(H)$.
\item $tw(Torso(G,C \cup \{s,t\})) \leq g(k,r)$. 
\end{enumerate}
\end{claim} 

Assume the claim is true.
By the last statement of Lemma \ref{smallprop}
$Torso(G, C \cup \{s,t\})=Torso(H, C \cup \{s,t\})$
Also, it is straightforward to check that a subset of $V(H)$
is an $s,t$-separator of $H$ if and only if it is an
$s,t$-separator of $G$. 
It follows that the algorithms specified in the claim in fact
returns the output requested by the theorem.
It thus remains to prove the claim.

The claim is similar to Theorem \ref{theorem215} except for the extra qualifier that $C$ must
be a subset of $V(H)$.  In order to 'get around' this qualifier, we define a graph $G^*$ where
the set of all minimal $s,t$-separators is guaranteed to be a subset of $V(H)$.
The graph $G^*$ is obtained from $G$ by introducing $k \cdot r+1$ copies of each vertex of $E(H)$. 
Specifically, $G^*[V(H)]=G[V(H)]$, for each $v \in V(H)$ and each copy $u'$ of a vertex $u \in E(H)$,
$v$ is adjacent with $u'$ if and only if $v$ is adjacent with $u$ in $G$, and finally for every two distinct
vertices $u_1,u_2 \in E(H)$, each copy $u'_1$ of $u_1$ is adjacent to each copy $u'_2$ of $u_2$ if
and only if $u_1$ is adjacent to $u_2$ in $G$. 

Let $X$ be a minimal $s,t$-separator of $G^*$ of size at most $k \cdot r$.
Observe that $X \subseteq V(H)$. Indeed, otherwise, $X$ contains a copy $u'$ of a vertex
$u \in E(H)$. Since there are $k \cdot r+1$ such copies there must be a copy $u''$ of $u$ that does not belong to 
$X$. By the minimality of $X$, $G^*$ a path $P$ from $s$ to $t$ containing $u'$ and not containing any vertex
of $X \setminus \{u'\}$. Replace in $P$ $u'$ by $u''$. As a result we obtain a walk from $s$ to $t$ that does not
intersect with $X$ as walk contains a path, this contradicts $X$ being an $s,t$-separator of $G^*$. 

Apply Theorem \ref{theorem215} to graph $G^*$ 
instead of $G$ and let $C$  be the resulting set.
To prove the claim,
it remains to establish the following two statements.
\begin{enumerate}
\item $Torso(G^*,C)=Torso(G,C)$.
\item $C$ includes all the minimal $s,t$-separators $X \subseteq V(H)$ of $G$ having size at most $p$.  
\end{enumerate}

For the first statement, as $Torso(G,C)$ is clearly a subgraph of $Torso(G^*,C)$, 
let $u,v$ be two adjacent vertices of $Torso(G^*,C)$.
This means that $G^*$ has a $u,v$-path $P$ where all the intermediate vertices do not occur in
$C$. Replace each copy $w'$ of a vertex of $w \in E(H)$ occurring on $P$ by $w$ itself.
As a result, we obtain a $u,v$-walk of $G$ where the intermediate vertices do not occur in $C$,
clearly the walk can be transformed into a path with the same property.

For the second statement, it is enough to show that if $X \subseteq V(H)$ is an $s,t$-separator
of $G$, the same is true regarding $G^*$. Indeed, otherwise, arguing as in the previous
paragraph, an $s,t$-path of $G^* \setminus X$ can be transformed into an $s,t$-walk of $G$
containing such a path, in contradiction to the definition of $X$.
$\blacksquare$

Theorem \ref{upgrade} harnesses the Treewidth Reduction Theorem
to obtain a graph of bounded treewidth having the same minimal $s,t$-separators
of size at most $k \cdot r$ as $H$. One important aspect is, however, missing:
we no longer know the edge cover number of the separators without having a look at $H$.
This is exactly what the extended torso is needed for!

\begin{lemma} \label{movetoextended}
Let $H$ be a hypergraph, let $U \subseteq V(H)$ , $s,t \in U$, and $S \subseteq U \setminus \{s,t\}$.
Then $S$ is a $s,t$-separator of $H$ with edge cover number at most $k$ if and only
if $S$ is an $s,t$-separator of $Torso^*(H,U)$ that is dominated by at most $k$ vertices of
$Torso^*(H,U) \setminus Torso(H,U)$. 
\end{lemma}

{\bf Proof.}
It is immediate from definition of the extended torso that
$\rho_H(S) \leq k$ if and only if in $Torso^*(H,U)$, $S$ can be dominated
by at most $k$ vertices of $Torso^*(H,U) \setminus Torso(H,U)$: the covering edges
correspond to the dominating vertices. 
By the second statement of Lemma \ref{smallprop},
$S$ is an $s,t$-separator of $Torso^*(H,U)$ if and only if
$S$ is an $s,t$-separator of $Torso(H,U)$. It remains to prove that
$S$ is an $s,t$-separator of $Torso(H,U)$ if and only if $S$ is an $s,t$-separator of $H$. 

Let $P$ be an $s,t$-path of $Torso(H,U)$ avoiding $S$. 
Then $P$ can be transformed to an $s,t$-path in $H$ by possibly introducing
some vertices outside of $U$. Since $S \subseteq U$, the resulting $s,t$-path in $H$
also avoids $S$.
Conversely, an $s,t$ path of $H$ avoiding $S$ can be transformed into a 
path in $Torso(H,U)$ by removal of vertices outside $U$ thus clearly producing
a path avoiding $S$. 
$\blacksquare$

Thus combination of Theorem \ref{upgrade}, Lemma \ref{movetoextended}, and the 
third statement of Lemma \ref{smallprop} reduce testing existence of a separator of a bounded
edge cover number in a bounded rank hypergraph to testing existence of a constrained separator
in a graph of bounded treewidth.  To enable the use of Courcelle's 
theorem, we need to demonstrate that the required separation property is \textsc{mso} definable. 
This is established in the next lemma. 

\begin{lemma} \label{lemcourc}
Let $G=(V,E)$ be a graph, let $U_1,U_2$ be two disjoint sets with $U_1 \cup U_2=V$,
let $v_1,v_2$ be two distinct vertices of $U_1$.
Let $k \geq 1$ be an integer. 
Then there is an \textsc{mso} formula $\varphi_k$ of length linear in $k$ 
that satisfies the model $(V,E,U_1,U_1,v_1,v_2)$ if and only if 
there is a $v_1,v_2$-separator $X \subseteq U_1$
that can be dominated by at most $k$ vertices of $U_2$.  
\end{lemma}

{\bf Proof.}
The formula is
$\exists X \exists D \exists R_1 \exists R_2  (X \subseteq U_1) \wedge (D \subseteq U_2) \wedge AtMost(D,k) \wedge Dominates(D,X) 
  \wedge R_1(v_1) \wedge R_2(v_2) \wedge (\neg R_1(v_2)) \wedge (\neg R_2(v_1))\wedge Comp(R_1,X) \wedge Comp(R_2,X)$. 

In this formula $AtMost(D,k)$ means that $|D| \leq k$.
In \textsc{mso} terms in can be expressed as
$\exists u_1 \dots \exists u_k \forall u D(u) \rightarrow (u=u_1 \vee \dots \vee u=u_k)$. 

$Dominates(D,X)$ means that $D$ dominates $X$. In terms of \textsc{mso} it can be expressed as
$\forall u \exists v X(u) \rightarrow (D(v) \wedge E(u,v))$.

Finally, $Comp(R,X)$ means that in $G \setminus X$ there is no path between $R$
and $V(G) \setminus (X \cup R)$. In terms of \textsc{mso} it is defined as 
$\forall u \forall v (R(u) \wedge E(u,v)) \rightarrow (R(v) \vee X(v))$.

The correctness of the formula 
is evident from the description.
$\blacksquare$

{\bf Proof of Theorem \ref{mainghw21}.}
The resulting algorithm consists of the following three steps.

\begin{enumerate}
\item Run the algorithm as specified by Theorem \ref{upgrade}.
As a result we obtain a set $C$  including all minimal $s,t$-separators of $H$
of edge cover at most $k$.  Let $G_0=Torso(H,C \cup \{s,t\})$. 
It is guaranteed that $tw(G_0) \leq g(k,r)$ where $g$ is the function
as in Theorem \ref{upgrade}. 
\item Let $G_1=Torso^*(H, C \cup \{s,t\})$ 
By the third item of Lemma \ref{smallprop}, $tw(G_1) \leq tw(G_0)+1 \leq g(k,r)+1$
\item If $G_1$ has an $s,t$-separator $S \subseteq V(G_0)$ dominated by at most $k$
vertices of $V(G_1) \setminus V(G_0)$ then return $true$ otherwise return $false$. 
In order to do the testing first apply a linear time algorithm \cite{BodTWD} for computation
of bounded tree decomposition of $G_1$. Then apply Courcelle's theorem
to test $(G_1,V(G_0),V(G_1) \setminus V(G_0), s,t) \models \varphi_k$
where $\varphi_k$ is the \textsc{mso} formula as specified in Lemma \ref{lemcourc}
\end{enumerate}

It is clear from the description that the above algorithm is \textsc{fpt} 
in $k$ and $r$. It remains to prove correctness of the algorithm. 
Let $S$ be an $s,t$-separator of $H$ of edge cover at most $k$.
We can assume w,l.o.g. that $S$ is a minimal separator for otherwise we can take its minimal
subset. This means that, by construction, $S \subseteq C$.   
By Lemma \ref{movetoextended}, $S$ is an $s,t$-separator of $G_1$.
Hence, by Lemma \ref{lemcourc}, the algorithm returns $true$. 
Conversely, suppose the algorithm returns $true$. By Lemma \ref{lemcourc},
there is a set $S \subseteq V(G_0)$ that separates $s$ and $t$ in $G_1$ and is dominated
by at most $k$ vertices of $V(G_1) \setminus V(G_2)$. 
By Lemma \ref{movetoextended}, this means that $S$ is at $s,t$-separator of $H$
of edge cover number at most $k$.

$\blacksquare$

\subsection{GHW computation}

\begin{theorem}

There is a constant $c$ so that for 
every fixed $r$ and $k$ the following two statements hold.
\begin{enumerate}
\item Let $H$ be a hypergraph  of rank at most $r$.
Then there is an $O((|V(H)|+|E(H)|^c \cdot |V(H)| \rho(H)^2)$ algorithm
then returns a tree decomposition of $H$ of \textsc{ghw} at most $4k$ or 'NO'.
In the latter case, it is guaranteed that $ghw(H)>k$.
\item Let $H$ be a hypergraph  of rank at most $r$.
Then there is an $O((|V(H)|+|E(H)| ^c\cdot \rho(H)^2)$ algorithm
then returns a tree decomposition of $H$ of \textsc{ghw} at most $6k$ or 'NO'.
In the latter case, it is guaranteed that $ghw(H)>k$.
\end{enumerate}

\end{theorem} 

{\bf Proof.}
By Theorem \ref{mainghw22}, the balanced separator
can be computed in \textsc{fpt} time parameterized by $k$ and $r$.
The sets to which the function $IsCover$ are of size $O(k\cdot r)$,
hence the testing can be done in a brute-force manner.
With these auxiliary functions, the runtime of a single
application of $ApproxGHW$ is \textsc{fpt} parameterized by $k$ and $r$. 
The theorem is now immediate from Theorem \ref{ghwpattern}.

$\blacksquare$

\section{Computation of fractional hypertree-wdith}
The algorithm for computation of \textsc{fhw} for hypergraphs of bounded
rank follows the same approach as the algorithm for \textsc{ghw}.
The proofs of several statements are identical in both cases. 
In these cases, we refer to the proof of the corresponding earlier statement. 
As a result, the description in this section is much shorter than in the previous one.
Hence, we do not divide it into subsections. 

\begin{definition}
Let $H$ be a hypergraph, $S,X \subseteq V(H)$.
We say that $X$ is a balanced \emph{fractional} separator for $S$
if for each connected component $V'$ of $H \setminus X$,
$\rho^*(V' \cap S)< 2/3 \rho^*(S)$, 
\end{definition}

\begin{theorem} \label{balancedfrac}
Let $H$ be a hypergraph, let $k>0$ be a real number.
Let $S \subseteq V(H)$ be such that $\rho^*(S)>3k$. 
Suppose that $fhw(H) \leq k$. 
Then $S$ has a balanced fractional separator $X$ with $\rho^*(X) \leq k$.
\end{theorem}

{\bf Proof.} 
Analogous to the proof 
of Theorem \ref{balancedsep} with the $\rho^*$  parameter used instead
of $\rho$.
$\blacksquare$

\begin{definition}
Let $BalSepFR(H,S,k)$ be a function testing existence of a balanced fractional separator $X$
of $S$ with $\rho^*(X) \leq k$.
Let $IsCoverFR(H,S,k)$ be a function testing whether the fractional edge cover of $S$ is at most $k$. 
We define the procedure $ApproxFHW(H,S,k,p)$ analogously to 
$ApproxGHW(H,S,k,p)$ with $BalSepFR$ used instead $BalSep$, $IsCoverFR$ used instead
$IsCover$ and $\rho^*$ substituted instead of $\rho$.   
\end{definition}

\begin{lemma} \label{frac111}
Let $H$ be a hypergraph and $S \subseteq V(H)$. 
Suppose that $\rho^*(S)>3k$. Let $X$ be a balanced separator of $S$
with $\rho^*(X) \leq k$. Then $H \setminus X$ has at least two connected components. 
\end{lemma}

{\bf Proof.}
Analogous to the first statement of Lemma \ref{lem111}
with $\rho^*$ used instead of $\rho$.
$\blacksquare$

\begin{lemma} \label{frac114}
Let $H$ be a connected hypergraph and $S \subseteq V(H)$. 
Let $k>0$ be an integer. Then $ApproxFHW(H,S,k,3k)$ 
invokes the function $ApproxGHW$ at most $|V(H)| \cdot \rho^*(V(H))^2$
times.  
\end{lemma}

{\bf Proof.}
Analogous to Lemma \ref{lem114} with $\rho^*$ used instead of $\rho$. 

\begin{theorem} \label{fhwpattern}
Let ${\bf H}$ be a class of hypergraphs. 
Suppose that for each $H \in {\bf H}$ a single recursive application
of $ApproxFHW$ takes time $x(H,k)$.
Let $H \in {\bf H}$ be a connected hypergraph. 
Then $ApproxFHW(H, \emptyset,k,3k)$ returns a tree decomposition of $H$
of \textsc{fhw}  less than $4k+1$ or 'NO'. In the latter case, it is guaranteed that 
$fhw(H)>k$. The runtime of the algorithm is 
$O(x(H,k) \cdot |V(H)| \cdot \rho(H)^2)$.
\end{theorem}

{\bf Proof.}
The proof of the theorem \emph{largely} follows the proof of Theorem \ref{ghwpattern}. 
However,  unlike the previous three statements, the modification here is more nuanced,
hence we consider it in greater detail. 
First of all,  the runtime is immediate from Lemma \ref{frac114}.
Next, note that, by construction, if $ApproxFHW$ attempts to construct a balanced separator
for  a set $S$ then $\rho^*(S)>3k$. Therefore,
it is immediate from Theorem \ref{balancedfrac} that if $ApproxFHW$ returns 'NO'
then the \textsc{fhw} of the input hypergraph is greater than $k$.
Next, we observe that if $ApproxFHW$ does not return 'NO' then its output is a 
tree decomposition of the input hypergraph $H$, the reasoning being analogous 
to the corresponding reasoning for $ApproxGHW$.

Finally, we observe that the \textsc{fhw} of the resulting decomposition is less than $4k+1$
and this is where the reasoning deviates from the corresponding reasoning for $ApproxGHW$. 
We establish this for $ApproxFHW(H,S,k,3k)$ provided that $\rho^*(S)<3k+1$.
If $ApproxFHW(H,S,k,3k)$ does not apply itself recursively, this is immediate by construction.

Otherwise, let $S'$ be the second parameter of the input 
of the recursive application.
Let $S^*$ be obtained from $S'$ as a result of the first loop. 
If $\rho^*(S')>3k$ then $S^*=S'$ since no vertices are added. 
Otherwise, $S^*$ is obtained by adding one vertex
to a set with fractional edge cover number smaller than $3k$. Clearly, the fractional edge cover number
of the resulting set cannot be larger than $3k+1$. It follows that $2/3\rho^*(S^*)<2k+1$.
Therefore, for each recursive application of $ApproxFHW$, the fractional edge cover number of
the set parameter is less than $2k+1+k=3k+1$. By the induction assumption, the \textsc{fhw} of the tree decomposition
returned by these applications is less than $4k+1$. It remains to observe that, by construction, the upper
bound holds for the bags formed within $ApproxFHW(H,S^*,k,3k)$ without applying recursion.
$\blacksquare$

Now we demonstrate that for hypergraphs of rank at most $r$
testing existence of a fractional balanced separator of fractional edge cover number at most $k$
is \textsc{fpt} parameterized by $\lceil k \rceil$ and $r$. 
Our first step is to observe that if $\rho^*(U) \leq k$ then $|U| \leq k \cdot r$.
In the integral case, this was immediate from the rank restriction.
In the fractional case, we need to a simple double counting argument.

\begin{lemma} \label{fracsize1}
Let $H$ be a hypergraph of rank at most $r$.
Let $U \subseteq V(H)$  and suppose that $\rho^*(U) \leq k$.
Then $|U| \leq k\cdot r$.
\end{lemma}

{\bf Proof.}
Consider a bipartite graph $G$ where the parts are $U$ and
the set $A$ of hyperedges of $H$ having a non-empty intersection
with $U$. For each $u \in U$ and $e \in A$, there is an edge between $u$ and $e$
if and only if $u \in e$.

$\rho^*(U) \leq k$ is equivalent to assigning each $e \in A$, a weight $w(e) \in [0,1]$
so that for each $u \in U$, $\sum_{u \in e} w(e) \geq 1$ and $\sum_{e \in A} w(e) \leq k$.
Assign to each edge $(u,e)$ the weight $w(e)$. Because of the rank assumption, the degree of 
each $e \in A$ is at most $r$ and hence 
the total weight of the edges of $G$ is at most $k \cdot r$.

Let $u_1, \dots, u_q$  be the vertices of  $U$.
Let $A_1, \dots, A_q$ be the partition of edges of $G$ so that the edges
of $A_i$ are incident to $u_i$.
As $\sum_{i=1}^q w(A_i) \leq k \cdot r$
and for each $A_i$, $w(A_i) \geq 1$, $q \leq k \cdot r$ by the pigeonhole
principle.
$\blacksquare$

The approach we used for the integral case was reducing computation of balanced separator to 
the computation of an ordinary separator between two vertices. 
This approach was enabled by Lemma \ref{lemmul3}.
In the fractional case, the lemma does not apply. However, we prove the following relaxed version 
allowing to reduce the balanced separator computation to computation of a $q$-way
cut where $q \in \{2,3\}$. 

\begin{lemma} \label{lemfrac3}
Let $H$ be a hypergraph, $S \subseteq V(H)$ with $\rho^*(S)>3k$.
Let $X$ be a fractional balanced separator for $S$ with $\rho^*(X) \leq k$. 
Then $H \setminus X$ is the union of $q \in \{2,3\}$ 
vertex disjoint subgraphs $H_1, \dots, H_q$
such that for each $1 \leq i \leq q$,
$\rho^*(V(H_i) \cap S)<2/3\rho^*(S)$. 
\end{lemma}

{\bf Proof.}
Let $V_1, \dots, V_r$ be the connected components
of $H \setminus X$. 
By Lemma \ref{frac111}, $r>1$. 
If $r \leq 3$, we are done. Otherwise, 
assume that there is $1 \leq i \leq r$
such that $\rho^*(V_i \cap S)>1/3\rho^*(S)$. 
W.l.o.g., we may assume that $i=1$. 
By the reasoning analogous to the proof of Corollary \ref{decomprho},
$\rho^*(V_1 \cup \dots \cup V_r)=\sum_{i=1}^r \rho^*(V_i)$,
we conclude that $\rho^*(S \cap (V_2 \cup \dots \cup V_r))<2/3 \rho^*(S)$.
Hence, we set $H_1=H[V_1]$ and $H_2=H[V_2 \cup \dots \cup V_r]$.
Note that if for each $1 \leq i \leq r$, $\rho^*(V_i \cap S)=1/3\rho^*(S)$
then $r \in \{2,3\}$, a possibility we have already excluded.
Therefore, it remains to assume that for each $1 \leq i \leq r$
, $\rho^*(V_i \cap S) \leq 1/3 \rho^*(S)$ and, for at least one $i$,
$\rho^*(S \cap V_i)<1/3 \rho^*(S)$. Assume w.l.o.g. that $i=1$.
Then apply the above argument inductively to 
$V_1 \cup V_2, V_3, \dots, V_r$.
$\blacksquare$  

Now we define a hypergraph obtained by contraction of several
disjoint sets of vertices.  Even though we contract at most $3$ sets,
we provide a definition for an arbitrary number of contracted sets. 

\begin{definition}
Let $H$ be a hypergraph.
Let $q \geq 1$ be an integer.
Let $S_1, \dots, S_q$ be mutually disjoint non-empty subsets of $V(H)$.
Let $\pi=(S_1 \leftarrow s_1, \dots S_q \leftarrow s_q)$ where $s_1, \dots, s_q$
are $q$ new vertices not present in $H$. 
We define $H_{\pi}$ recursively as follows. 
If $q=1$ then $H_{\pi}=H_{S_1 \leftarrow s_1}$.
Otherwise, let $\pi'={S_2 \leftarrow s_2, \dots S_q \leftarrow s_q}$
Then $H_{\pi}=[H_{S_1 \leftarrow s_1}]_{\pi'}$. 

Assume that $S_1, \dots S_q$ as above are subsets of a set $S \subseteq V(H)$. 
Then hypergraph $H^S_{\pi}$ is obtained from $H_{\pi}$ 
by connecting each vertex $u \in S \setminus (S_1 \cup \dots \cup S_q)$ by
an edge to each $s_1, \dots, s_q$.
\end{definition}

\begin{lemma} \label{frac22}
Let $H$ be a hypergraph without isolated vertices, and let $k>0$ be a real.
Let $S \subseteq V(H)$ be such that $\rho^*(S)>3k$. 
Then $X$ s a balanced separator of $X$ with $\rho^*(X) \leq k$
if and only if there are disjoint subsets $S_1, \dots S_q$ ($q \in \{2,3\}$)
with $\rho^*(S_i)<2/3\rho^*(S)$ for each $1 \leq i \leq q$ so that the following statements
hold.
\begin{enumerate}
\item $X$ is an $s_1, \dots, s_q$-cut of $H^*=H^S_{\pi}$ where $\pi=(S_1 \leftarrow s_1, \dots S_q \leftarrow s_q)$.
\item $\rho_{H^*}(X) \leq k$. 
\end{enumerate}
\end{lemma}

{\bf Proof.}
The proof largely follows the reasoning of Lemma \ref{lem22}.
Let $X$ be a balanced separator of $S$ with fractional edge cover number at most $k$. 
Let $H_1, \dots, H_q$ ($q \in \{2,3\}$) be as guaranteed by Lemma \ref{lemfrac3}. 
For each $1 \leq i \leq q$, let $S_i=S \cap V(H_i)$. 
It follows that $S \setminus (S_1 \cup \dots \cup S_q) \subseteq X$
and that $X$ is a multiway cut for $s_1,\dots, s_q$ in $H^*$ as per definition.
Moreover, it is not hard to see that $\rho^*_{H^*}(X) \leq k$. 
Indeed, for each hyperedge of $H^*$ assign the weight which is the sum of weight
of its preimages (capping the resulting weight by $1$ if needed).

Conversely, suppose that there are disjoint subsets
$S_1, \dots, S_q$ with $\rho^*_{H}(S_i) \leq 2/3 \rho S$ such that
in the corresponding graph $H^*$ there is a multiway cut
$X$ with $\rho^*_{H^*}(X) \leq k$. 
Observe that $\rho^*_H(X) \leq k$. 
Indeed, let $\gamma^*: E(H^*) \rightarrow [0,1]$ be a function witnessing
a fractional edge cover of $X$ of the smallest possible weight. 
The edges assigned with non-zero weights can be seen as the union of two
disjoint sets $E_1$ and $E_2$ where $E_1 \subseteq E(H_{\pi})$ and
$E_2 \subseteq E(H^*) \setminus E(H_{\pi})$. 
For each $e^* \in E_1$, let $e$ be an arbitrary hyperedge of $H$ whose image in $H_{\pi}$ is $e^*$.
Assign $e$ with weight $\gamma_1(e)=\gamma(e^*)$. For all the edges $e$ outside $E_1$,, let $\gamma_1(e)=0$.   
For each edge $e^* \in E_2$, note that this edge contains precisely one vertex $u^*$ of $X$
(this is a vertex of $S \setminus (S_1 \cup \dots \cup S_q)$). 
Let $y$ be a mapping that maps each hyperedge $e^* \in E_2$ to an arbitrary hyperedge $e \in E(H)$ containing $u^*$. 
For each edge $e \in y(E_2)$, let $\gamma_2(e)$ be the sum of $\gamma^*$-weights of al the edges in the preimage 
of $e$. For the rest of the edges of $E(H)$, their $\gamma_2$-weights are zeroes.
Now, for each $e \in E(H)$, let $\gamma(e)=\gamma_1(e)+\gamma_2(e)$ (capped at $1$).
A direct inspection shows that the total weight of $\gamma$ is at most the weight of $\gamma^*$
and each vertex of $X$ is covered by $\gamma$.

It remains to demonstrate that $X$ is a fractional balanced separator for $S$.
Let $V'$ be a connected component of $H \setminus X$. By definition of $X$ and a direct
inspection, it is not hard to observe that there is $1 \leq i \leq q$
such that $S \cap V' \subseteq S_i$. We conclude that $\rho^*_{H}(S \cap V')<2/3 \rho^*_H(S)$
as required.
$\blacksquare$

\begin{theorem}  \label{fracsep}
Let $H$ be a hypergraph of rank at most $r$.
Let $k>0$ be a real.
Let $q \in \{2,3\}$. 
Let $s_1, \dots, s_q$ be $q$ distinct vertices of $H$.
Then there is an \textsc{fpt} algorithm parameterized by $\lceil k \rceil$ and $r$
that tests whether $H$ has a multiway cut $X$ for $s_1, \dots, s_q$ with 
$\rho^*(X) \leq k$.  
\end{theorem}

Assuming correctness of Theorem \ref{fracsep},
we are now in a position to establish constant ratio approximability of \textsc{fhw} 
for hypergraphs of bounded rank.

\begin{theorem} \label{mainfhw}
Let $r>0$ be an integer and let $k>0$ be real.
Then there is an \textsc{fpt} algorithm parameterized by $\lceil k \rceil$ and $r$
whose input is a hypergraph $H$ of rank at most $r$.
The algorithm either returns a tree decomposition of $H$ of fhw
at most $4k+1$ or 'NO'. In the latter case it is guaranteed that 
$fhw(H)>k$.
\end{theorem}

{\bf Proof.}
Taking into account that finding an optimal fractional edge cover of a subset of $V(H)$
can be done in a polynomial time by linear programming techniques,
the theorem will immediately follow from Theorem \ref{fhwpattern} if we establish
fixed-parameter tractability of balanced separator testing. 

As we observed in the proof Theorem \ref{fhwpattern} when the considered algorithm
tests existence of a balanced separator for a set $S$, it is guaranteed that
$3k< \rho^*(S) \leq 3k+1$. It follows from Lemma \ref{fracsize1} that $|S|=O(k \cdot r)$.
Therefore, we can consider in a brute-force manner all disjoint subsets $S_1, \dots, S_q$
of $S$ ($q \in \{2,3\}$). For every family of such subsets, we test whether $\rho^*(S_i)<2/3 \rho^*(S)$
for each $1 \leq i \leq q$ (the testing can be done in a polynomial time using 
linear programming techniques). If this holds we create a graph
$H^*=H^S_{\pi}$ where $\pi=(S_1 \leftarrow s_1, \dots S_q \leftarrow s_q)$ and test
whether $H^*$ has a multiway cut for $s_1, \dots, s_q$ of fractional edge cover number at most $k$.
This testing is \textsc{fpt} in the considered parameters by Theorem \ref{fracsep}.
If for at least considered $H^*$ the testing is positive the algorithm returns $True$.
Otherwise, it returns $False$.  
The correctness of balanced separator testing is immediate from Lemma \ref{frac22}.
$\blacksquare$

It now remains to prove Theorem \ref{fracsep}.

The following theorem is a special case of the Treewidth Reduction Theorem
(Theorem 2.15 of \cite{TALG}). 
\begin{theorem} \label{frac215}
Let $G=(V,E)$ be a graph, $s_1, \dots, s_q$ be $q$ distinct vertices ($q \in \{2,3\}$) and $p \geq 1$
be an integer. Then there is a function $h$ and an \textsc{fpt} algorithm paramterized by $p$
that computes a set $C$ 
such that $C$ is a superset of the union of all minimal $s_i,s_j$-separators of $G$ of size at most $p$ $1 \leq i \neq j \leq q$). 
Moreover, $tw(Torso(G,C \cup \{s_1, \dots, s_q\}) \leq h(p)$.
\end{theorem}

We harness the treewdith reduction theorem in the same way as we did for the integral case. 
The proof uses the argument analogous to the proof of Theorem \ref{upgrade} 
easily adapted to the multiway cut (rather than just a cut) setting. 

\begin{theorem} \label{fracupgrade}
Let $H$ be a hypergraph of rank at most $r$, $s_1, \dots, s_q$ be $q$ distinct vertices ($q \in \{2,3\}$) and $k>0$
be a real. 
There is a function $g$ and an \textsc{fpt} algorithm parameterized by $\lceil k \rceil$ and $r$
that returns a set  $C$ such that $C$ is a superset of the
union of all minimal $s_i,s_j$-separators of size at $k \cdot r$ ($1 \leq i \neq j \leq q$) and
$tw(Torso(H,C \cup \{s_1,\dots, s_q\})) \leq g(k,r)$.
\end{theorem}

Like in the integral case, our next step is to move from the torso to the extended torso
and to use Courcelle's theorem. However, in order to do this we need to demonstrate
that bounded fractional edge cover number is an \textsc{mso} definable property.
We do this below.

\begin{corollary} \label{fracsize2}
There is a function $h^*$ such that the following holds.  
Let ${\bf H}_{k,r}$ be the set of all hypergraphs $H$ of rank at most $r$ with 
$\rho^*(H) \leq k$. Then $|{\bf H}(k,r)| \leq h^*(\lceil k \rceil,r)$. 
\end{corollary}

{\bf Proof.}
It follows from Lemma \ref{fracsize1}
that ${\bf H}_{k,r}$ is a subset of hyeprgraphs of at most $k \cdot r$
vertices. Clearly, the number of such graphs is upper bounded by a function
depending on $k$ and $r$.
$\blacksquare$

\begin{lemma} \label{fracmsobound}
Let $r \geq 1$ be an integer and $k>0$ be a real. 
There is an \textsc{mso} predicate $AtMostFr(X,k,r)$ over models $(V,E,U_1,U_2)$
with $G=(V,E)$ being a graph, $U_1 \cup U_2=V$, and $U_1 \cap U_2=\emptyset$  
so that the following holds.

Let $H$ be a hypergraph of rank at most $r$, $U \subseteq V(H)$ and $X \subseteq U$. 
Let $U^*=V(Torso^*(H,U))$ and $E=E(Torso^*(H,U))$.
Then $(U^*,E,U^* \setminus U,U) \models AtMostFR(X,k,r)$ if and only if
$\rho^*_H(X) \leq k$. 
Furthermore the size of $AtMostFr(X,k,r)$ is upper bounded by a function of 
$\lceil k \rceil$ and $r$. 
\end{lemma}

{\bf Proof.}
In light of Lemma \ref{fracsize1},
the function can be represented as $\bigvee_{q=1}^{k \cdot r} AtMostFr(X,k,r,q)$
where the predicate with four parameters imposes an additional constraint
that $|X|=q$. 

Let ${\bf H}^q_{k,r}$ be the subset of ${\bf H}_{k,r}$ consisting of hyoergraphs with precisely 
$q$ vertices.
Fix a $1 \leq q \leq k \cdot r$ and let ${\bf H}^q_{k,r}=\{H_1, \dots, H_m\}$.
Suppose that $|X|=q$.
Then $AtMostFr(X,k,r,q)$ holds if and only if there is $1 \leq i \leq m$
such that $E(H_i) \subseteq E(H[X])$ (of course we need to set $V(H_i)=X$). 

Fix $1 \leq i \leq m$. 
Let $x_1, \dots, x_q$ be an arbitrary enumeration of the vertices of $X$
and let $e_1, \dots, e_b$ be an arbitrary enumeration of $E(H_i)$ 
as subsets of $x_1, \dots, x_q$.
Then $E(H_i) \subseteq E(H[X])$ if and only if there are
$y_1, \dots, y_b \in U^* \setminus U$ such that 
for each $1 \leq j \leq b$ and for each $1 \leq a \leq q$, $v_j$ is adjacent to $x_a$
in $Torso^*(H,U)$ if and only if $x_a \in e_j$. 

Based on the reasoning in the previous paragraph, let us
define a predicate $Subgraph_{H_i}(x_1, \dots, x_q)$
as $\exists y_1, \dots, y_b \bigwedge_{j=1}^b (\bigwedge_{x_a \in e_j} E(x_a,y_j)  \wedge \bigwedge_{x_a \notin e_j} \neg E(x_a,y_j))$.
Taking into account the definition of an extended
torso, it is not hard to see that provided $x_1, \dots, x_q$ are distinct elements of $U$,
$Subgraph_{H_i}(x_1, \dots, x_q)$ is true if and only if $E(H_i) \subseteq E(H[X])$. 

With this in mind $AtMostFr(X,k,r,q)$ by asserting the existence of 
vertices $x_1, \dots, x_q$ that are pairwise distinct and each vertex of $X$ is one of them
and, finally that $Subgraph_{H_i}(x_1, \dots, x_q)$ 
is true for at least one $1 \leq i \leq m$. In terms of the \textsc{mso}, this is expressed
as follows.

$\exists x_1, \dots, x_q IsSet(X,x_1, \dots, x_q) 
\wedge \bigvee_{1 \leq i \leq m} Subgraph_{H_i}(x_1, \dots, x_q)$.
where $IsSet(X,x_1, \dots, x_q)$ is defined as 
$\bigwedge_{1 \leq c \neq d \leq q} (x_c \neq x_d) \wedge (\forall x \in X \bigvee_{1 \leq c \leq q} x=x_c)$

The required upper bound on the size of the resulting formula follows from a direct inspection.
$\blacksquare$ 

It remains to define an \textsc{mso} formula for testing whether the given
set of vertices is a multiway cut. 

\begin{lemma} \label{ismultiway}
There is an \textsc{mso} predicate $IsMultiway(X,S)$ over models $(V,E,U_1,U_2)$
with $G=(V,E)$ being a graph $U_1 \cup U_2=V$ and $U_1 \cap U_2=\emptyset$ so that 
$(V,E,U_1,U_2) \models IsMultiway(X,S)$ if and only if $X \subseteq U_1$
and $X$ is a multiway cut for $S$. 
\end{lemma}

{\bf Proof.}
The formula is $(X \subseteq U_1) \wedge \forall s \in S \exists R ( Disconnected(R,X) \wedge R(s) \wedge 
\forall s' \in S (s' \neq s \rightarrow \neg R(S')))$
where $Disconnected(R,X)$ means that $R$ is a union of connected components of $G \setminus X$. 
In particular, $Disconnected(R,X)$ may be encoded as follows.
$\forall u \in R \forall v E(u,v) \rightarrow (R(v) \vee X(v))$.
The correctness of the formula is verifiable by a direct inspection.
$\blacksquare$

{\bf Proof of Theorem \ref{fracsep}.}
The resulting algorithm consists of the following three steps.

\begin{enumerate}
\item Run the algorithm as specified by Theorem \ref{fracupgrade}.
As a result we obtain a set $C$  including all minimal $s_i,s_j$-separators of $H$
$(1 \leq i \neq j \leq q$)
of size at most $k \cdot r$ and also a graph $G_0=Torso(H,C \cup \{s_1, \dots, s_q\})$. 
Moreover, it is guaranteed that $tw(G_0) \leq g(k,r)$ where $g$ is the function
as in Theorem \ref{fracupgrade}. 

\item Obtain $G_1=Torso^*(H, C \cup \{s_1,\dots, s_q\})$ 
By the third item of Lemma \ref{smallprop}, $tw(G_1) \leq tw(G_0)+1 \leq g(k,r)+1$
\item 
Apply an \textsc{fpt} algorithm for computing a bounded tree decomposition of $G_1$. 

Then test $(G_1,V(G_0),V(G_1) \setminus V(G_0), s_1, \dots, s_q) \models \exists X  IsMultiway(X, \{s_1, \dots,s_q\}) \wedge AtMostFr(X,k,r)$
Return the outcome of the testing. By Courcelle's theorem and Lemma \ref{fracmsobound},
the testing can be done in \textsc{fpt} time parameterized by $\lceil k \rceil$ and $r$.  
\end{enumerate}

It remains to establish the correctness of the algorithm. 
Suppose that there is $X \subseteq V(H)$ that is an $s_1, \dots s_q$-cut of $H$ and $\rho^*_H(X) \leq k$. 
We may assume w.l.o.g. that $X$ is minimal subject to this property.
By Lemma \ref{fracsize1}, $|X| \leq k \cdot r$. 
Let $u \in X$. Then there are $1 \leq i \neq j \leq q$ such $u$ belong
to a minimal $s_i,s_j$-separator of size at most $k \cdot r$. 
Indeed, for every such $i,j$ we can identify $X_{i,j} \subseteq X$
so that $X_{i,j}$ is a minimal $s_i,s_j$-cut of $H$. Clearly, the union of all $X_{i,j}$ is a multiway
cut for $s_1, \dots, s_q$ and hence equals $X$ due to its minimality. Hence $u$ must belong to some
$X_{i,j}$. We conclude that $X \subseteq C$. By the reasoning analogous to the integral case, we conclude
that $X$ is a $s_1, \dots, s_q$-cut for $G_1$. 
Consequently, $IsMultiway(X, \{s_1, \dots,s_q\})$ is true on $(G_1,V(G_0),V(G_1) \setminus V(G_0), s_1, \dots, s_q)$
by Lemma \ref{ismultiway}. 
Also, by Lemma \ref{fracmsobound} and the definition of $X$ 
$AtMostFr(X,k,r)$ is true on $(G_1,V(G_0),V(G_1) \setminus V(G_0), s_1, \dots, s_q)$.
Consequently, the algorithm returns $True$.
Conversely, assume that the algorithm returns $True$.  
By Lemmas \ref{ismultiway} and \ref{fracmsobound} 
there is $X \subseteq V(G_0)$ that is a multiway cut and $\rho^*_H(X) \leq k$.
Arguing similarly to the integral case, we conclude that $X$ is a mutlway cut for $s_1, \dots, s_q$
in $H$.
$\blacksquare$

\section{Conclusion}
In this paper we proposed the first \textsc{fpt} algorithm for constant approximation of \textsc{ghw} and \textsc{fhw} of
hypergraphs of bounded rank. 
In this section, we identify two natural directions of further research.
The first direction is to attempt to reduce the approximation ratio and, ideally, to produce an exact algorithm.
The approach based on computation of \emph{typical sequences} \cite{BodKlokstwd,Kloksbook} is a standard 
methodology for accomplishing this task in the context of treewidth computation. 
This approach takes as input a tree decomposition of the given graph of treewidth larger than the optimal
but still upper-bounded by a function of the parameter and uses a sophisticated refinement to produce 
a tree decomposition of the smallest treewidth. 
We conjecture that this approach can be adapted to the \textsc{fpt} computation of \textsc{ghw} and \textsc{fhw} 
for hypergraphs of bounded rank to produce approximation 
algorithms with ratio at most $2$. Moreover, we conjecture that exact \textsc{fpt} algorithms can be obtained
if we take the max-degree as the extra parameter.

Another possible approach for design of exact algorithms for computation of hyper treewidth parameters is
a recently local improvement method \cite{Kortw2approx, Kortwprec}. It would be particularly interesting to try this approach
for hypergraphs of bounded rank but unbounded degree.     

The second research direction is design of \textsc{fpt} algorithms approximating hyper treewidth parameters
for hypergraphs of unbounded rank.
We will concentrate on \textsc{ghw}. 
Let us first discuss how such an algorithm can be implemented using the $ApproxGHW$ framework.
Recall that it is sufficient to produce \textsc{fpt} algorithms for the oracles. 
In fact, we do not need to care about the edge cover testing as it is \textsc{fpt} for quite a general class
of hypergraphs of bounded intersection. So, we concentrate on testing existence of a balanced separator. 

For a hypergraph of bounded rank we reduced the problem of computing a balanced separator for a set $S$ to the problem of computing
a cut between two vertices. Specifically, we considered in a brute-force manner all combinations $S_1,S_2,S_3$ of disjoint sets
with $S_1 \cup S_2 \cup S_3=S$ and, for combinations answering an additional easily checkable constraint, tested
existence of an $S_1,S_2$-separator $X$ of $H \setminus S_3$ with $\rho(X) \leq k$.
With hyperedges being of unbounded rank, the size of $S$ becomes unbounded and hence such a brute-force testing 
is no longer feasible. However, we can easily get around this obstacle by considering an edge cover $E=\{e_1, \dots e_q\}$ of $S$
with $q=O(k)$.  We consider all possible combinations of disjoint subsets $E_1,E_2,E_3$ of $E$ with $E_1 \cup E_2 \cup E_3=E$. 
For each such a combination, let $S_1=\bigcup E_1 \setminus (\bigcup E_2 \cup \bigcup E_3)$ and let
$S_2=\bigcup E_2 \setminus (\bigcup E_1 \cup \bigcup E_3)$. 
For those non-empty $S_1$ and $S_2$ whose edge cover number is less than $2/3 \rho(S)$, we test the existence 
of an $S_1,S_2$-cut of $H \setminus (S \setminus (S_1 \cup S_2))$. of edge cover number at most $k$.
It is not hard to see that if the $S_1,S_2$-cut testing is \textsc{fpt} in $k$, the whole procedure is guaranteed to return
a balanced separator of edge cover at most $2k$ 
provided existence of a balanced separator of edge cover at most $k$. This will increase 
the approximation ratio of the resulting \textsc{ghw} computation but it will still remain constant. 
Thus we conclude that a cornerstone for an \textsc{fpt} approximation of \textsc{ghw} is computation
of a cut with a small edge cover number. We formalize this conclusion in the form of an open question.

\begin{question} \label{quest1}
Let $c>1$ be a constant, $k>0$ be an integer. Is there an \textsc{fpt} algorithm parameterized by $k$
whose input is a graph $H$ whose incidence graph does not have $K_{c,c}$ as a subgraph and two vertices $s$ and $t$ of $H$
so that the algorithm tests existence of an $s,t$ cut with edge cover number a most $k$?

As a special case, does such an algorithm exist under assumption that $H$ is a linear hypergraph,
(the intersection size of any two hyperedges is at most $1$)? 

As an even more restricted special case, is it possible to obtain such an algorithm under assumption that
the max-degree of a vertex of $H$ is upper-bounded by a constant? 

\end{question}

We believe the above question to be of an independent interest. 
In particular, we believe that the resolution of this question  
will provide a significant new insight into the mathematics of graph
separation problems.

It is interesting to explore a possibility of adaptation of the algorithm
of Bodlaender \cite{BodTWD} for the purpose of \textsc{ghw} approximation.
One advantage of this approach is that it does not involve computation 
of balanced separators. Intuitively the adaptation looks possible with,
for example, considering vertices with small edge covers of their neighbourhoods
instead of vertices of a small degree. However, this approach involves another
inherent challenge. A main recursive step of the algorithm is computing a large
matching and contracting the edges of the matching. The treewidth is then computed for the resulting
graph equipped with a witnessing tree decomposition.
When the graph is 'decontracted', we obtain a tree decomposition for the original
graph by placing the contracted edge in all the bags where the image vertex of the edge occurs.
Consequently, the treewdth of the resulting decomposition may effectively grow twice.
Therefore, iterations of 'decontraction' must be alternated with iterations of \emph{refinement}.
A refinement procedure takes as input a tree decomposition of a graph $G$ of width $2k$
and produces a tree decomposition of width at most $k$ or reports that the treewidth of $G$
is greater than $k$.  Adaptation of the approach to \textsc{ghw} approximation will require 
to design such a refinement for \textsc{ghw} and we consider it the main challenge of the adaptation.
Below we formally state a related open question.

\begin{question} \label{quest2}
Are there a function $f$, a constant $c$ and and \textsc{fpt} algorithm parameterized by $k$
doing the following. The input of the algorithm is a graph $H$ of a bounded multiple intersection 
and a tree decomposition
of $H$ of \textsc{ghw} at most $f(k)$. The output is a tree decomposition of $H$ of width at
most $ck$ or 'NO'. In the latter case it is guaranteed that $ghw(H)>k$. 
As in Question \ref{quest1}, consider this question also for linear hypergraphs with a possible extra
assumption of a constant max-degree. 
\end{question} 

A challenging aspect of Question \ref{quest2} is that, due to a bag size being unbounded in terms of $k$,
all subsets of a bag cannot be greedily explored. One way to get around this obstacle is to explore all
unions of subsets of edges in an edge cover of a bag.

\section*{Acknowledgement}
I would like to thank G. Gottlob, M. Lanzinger, and R. Pichler for helping me to
understand the concepts, ideas, and algorithms related to hypertree width parameters.
I would like to thank T. Korhonen for letting me know about the result \cite{Domsetfptnonapprox}.


\end{document}